\documentclass[aps,prd,final,twocolumn,letterpaper,superscriptaddress,nofootinbib,notitlepage]{revtex4-1}

\usepackage{graphicx}                       
\usepackage[outdir=./]{epstopdf}
\usepackage{amsmath}
\usepackage{amssymb}
\usepackage{amsthm}
\usepackage{dsfont}       
\usepackage[usenames,dvipsnames,table]{xcolor}
\usepackage[format=plain, font=small, labelfont=bf, justification = raggedright]{caption}
\usepackage{array}  
\usepackage{overpic}
\usepackage{placeins} 
\usepackage{fancyhdr}
\usepackage{colortbl}
\usepackage{multirow} 
\usepackage{hyperref} 
\hypersetup{colorlinks = true,linkcolor = blue, breaklinks = true}

\newcommand{\eq}[1]{(\ref{#1})}
\newcommand{\Eq}[1]{Eq.~\eq{#1}}
\newcommand{\Eqs}[1]{Eqs.~\eq{#1}}
\newcommand{\Fig}[1]{Fig.~\ref{#1}}
\newcommand{\Sec}[1]{Sec.~\ref{#1}}

\renewcommand{\Ref}[1]{Ref.~\cite{#1}}

\newcommand{\Refs}[1]{Refs.~\cite{#1}}
\newcommand{\App}[1]{Appendix~\ref{#1}}

\newcommand{\mc}[1]{\mathcal{#1}}

\newcommand{\msf}[1]{\mathsf{#1}}

\newcommand{\bra}[1]{\langle#1 |}
\newcommand{\ket}[1]{|#1 \rangle}
\newcommand{\braket}[2]{\langle#1 |  #2 \rangle}

\newcommand{\oper}[1]{\smash{\hat{#1}}}
\newcommand{\unit}[1]{\smash{\check{#1}}}
\newcommand{\fourier}[1]{\smash{\widetilde{#1}}}

\newcommand{\pd}[1]{\partial_{#1}}

\newcommand{\dd}{\mathrm{d}}

\DeclareMathOperator{\airyA}{Ai}
\DeclareMathOperator{\airyB}{Bi}
\DeclareMathOperator{\pearcey}{Pe}

\newcommand{\Vect}[1]{{\boldsymbol{\rm #1}}}
\newcommand{\VectOp}[1]{\oper{\Vect{#1}}}

\newcommand{\Mat}[1]{\msf{#1}}
\newcommand{\IMat}[1]{\Mat{I}_{#1}}
\newcommand{\OMat}[1]{\Mat{0}_{#1}}

\newcommand{\Tr}{\text{tr}}

\newcommand{\Symb}[1]{\mc{#1}}

\newcommand{\IdentOp}{\oper{\mathds{1}}}

\newcommand{\cont}[1]{\mc{C}_{#1}}

\newcommand{\aber}{a}

\newcommand{\nullFrac}{\vphantom{\frac{}{}}}

\newcommand{\Stroke}[1]{\text{\ooalign{ $#1$\cr \hidewidth\raise.225ex \hbox{$-\mkern.5mu$}\cr}}}

\begin{document}
\setlength{\parskip}{0pt}
\setlength{\belowcaptionskip}{0pt}


\title{Metaplectic geometrical optics for modeling caustics in uniform and nonuniform media}
\author{N. A. Lopez}
\affiliation{Department of Astrophysical Sciences, Princeton University, Princeton, New Jersey 08544, USA}
\author{I. Y. Dodin}
\affiliation{Department of Astrophysical Sciences, Princeton University, Princeton, New Jersey 08544, USA}
\affiliation{Princeton Plasma Physics Laboratory, Princeton, New Jersey 08543, USA}

\begin{abstract}

As an approximate theory that is highly regarded for its computational efficiency, geometrical optics (GO) is widely used for modeling waves in various areas of physics. However, GO fails at caustics, which significantly limits its applicability. A new framework, called metaplectic geometrical optics (MGO), has recently been developed that allows caustics of certain types to be modeled accurately within the GO framework. Here, we extend MGO to the most general case. To illustrate our new theory, we also apply it to several sample problems, including calculations of two-dimensional wavefields near fold and cusp caustics. In contrast with traditional-GO solutions, the corresponding MGO solutions are finite everywhere and approximate well the true wavefield near these caustics.

\end{abstract}

\maketitle

\pagestyle{fancy}
\lhead{Lopez \& Dodin}
\rhead{Metaplectic geometrical optics}
\thispagestyle{empty}


\section{Introduction}
\label{sec:intro}

The propagation of waves in homogeneous and weakly inhomogeneous media is often described within the approximate theory known as geometrical optics (GO), or ray optics~\cite{Kravtsov90,Tracy14}. However, GO fails at so-called caustics, where it predicts spurious singularities of the wavefield. Loosely speaking, caustics are surfaces across which the number of rays arriving at a given point changes abruptly~\cite{Kravtsov93}. The general properties of such surfaces have long been known from catastrophe theory, which provides a classification wherein only a finite number of caustic types are possible for a given number of spatial dimensions~\cite{Berry76,Berry80b}. This result underlies modern research into caustics~\cite{Hobbs07,Borghi16,Zannotti17a,EspindolaRamos19}, as general properties of a given caustic type can be inferred by studying a particular case. Still, practical calculations continue to rely on directly solving wave equations~\cite{Wright09,Shiraiwa10,Myatt17}, which is computationally expensive. It would be advantageous to find a more efficient way to calculate these caustic structures. In particular, the question whether caustics can be modeled by somehow extending GO has been attracting attention for a long time.

One known solution to this problem is by Maslov~\cite{Maslov81}, who proposed to rotate the ray phase space occasionally by $\pi/2$ using the Fourier transform (FT) in one or more spatial variables. Such rotations can remove caustics and locally reinstate GO, but they are inconvenient for simulations because the rotation points have to be introduced \textit{ad~hoc}, requiring the simulations to be supervised. In \Refs{Lopez19a,Lopez20a}, we proposed a modification of Maslov's approach by replacing the FT with the metaplectic transform (MT), which might be familiar from other contexts such as aberration theory~\cite{GarciaBulle86} and paraxial optics~\cite{Bacry81,Sudarshan83,Mukunda83}. Using the MT, one can transform the wavefield continually along the rays, both eliminating the singularities at caustics and allowing for fully automated and fast simulations. This new framework, which we call metaplectic geometrical optics (MGO), has already been successfully benchmarked on one-dimensional (1-D) problems~\cite{Lopez20a}. However, MGO as originally formulated in \Refs{Lopez19a,Lopez20a} can still yield singularities in certain situations; thus, further progress is needed.

Here, we report a more general version of MGO, where the last spurious singularities are removed at the expense of additional calculations. We then demonstrate the reformulated MGO analytically on three examples: a homogeneous plane wave, a fold caustic, and a cusp caustic. In Arnold's classification~\cite{Arnold75}, they correspond to the $A_1$, $A_2$, and $A_3$ caustic types, respectively. The first case has no caustic \textit{per~se}; it is considered only to present a basic tutorial on the MGO machinery. The two other cases show how MGO leads to solutions that, unlike traditional-GO solutions, are finite everywhere and approximate well the true wavefield near the caustics.

This paper is organized as follows. In \Sec{sec:MGO}, we review the basic equations of GO and MGO as described in \Ref{Lopez20a}. In \Sec{sec:MGO2}, we report a more general version of MGO. In \Sec{sec:ex}, we discuss three examples of the reformulated MGO, and we show analytically that MGO adequately approximates the wavefield in the entire space, even at caustics. In \Sec{sec:conclusion}, we summarize our main conclusions. Auxiliary calculations are presented in appendices and in the supplementary material.


\section{Overview of metaplectic geometrical optics}
\label{sec:MGO}

We start by briefly reviewing the MGO method as it was presented in \Ref{Lopez20a}. Let us consider a non-driven linear wave in a general linear medium. We do not assume any particular wave equation in this paper, as our theory is sufficiently general to handle any linear wave equation. That said, we shall restrict the analysis to scalar equations for simplicity. (Generalization to vector waves on non-Euclidean spaces is possible using the machinery presented in \Ref{Dodin19}.) 

Specifically, we assume that the wave is described by a scalar field $\psi$ and governed by an integral equation
\begin{equation}
    \int \dd \Vect{q}' \,
    D(\Vect{q}, \Vect{q}')
    \psi(\Vect{q}')
    = 0 .
    \label{eq:WAVEintegral}
\end{equation}

\noindent Here, $\Vect{q}$ is the coordinate on an $N$-D Euclidean (or pseudo-Euclidean) space, called $\Vect{q}$-space, and $D(\Vect{q}, \Vect{q}')$ is the dispersion kernel. In particular, differential wave equations (partial or ordinary) have dispersion kernels that consist of delta functions and their derivatives. For example, the Helmholtz equation has $D(\Vect{q},\Vect{q}') = \nabla'^2 \delta(\Vect{q}' - \Vect{q}) + n^2(\Vect{q}') \delta(\Vect{q}' - \Vect{q})$, where $\nabla'$ is the gradient with respect to $\Vect{q}'$ and $n(\Vect{q})$ is a spatially varying index of refraction. (See \Sec{sec:ex} for more examples.) More generally, the kernel $D$ can be a smooth function, as is the case for waves in warm plasma, for example. (A review of the general theory of linear dispersion can be found in \Ref{Dodin17d}.) It is not necessary for our purposes to specify this function; let us simply state that $D$ encodes all information about the linear medium where the wave propagates, whatever that medium may be.

Consider also $\Vect{p}$-space, which is the Fourier-dual of $\Vect{q}$-space. Then, $\Vect{q}$-space and $\Vect{p}$-space collectively define a $2N$-D phase space with coordinates $(\Vect{q}, \Vect{p})$. The dispersion kernel $D(\Vect{q}, \Vect{q}')$ can be associated with a function on phase space via the transformation
\begin{equation}
    \Symb{D}(\Vect{q}, \Vect{p})
    \doteq \int \dd \Vect{s} \, 
    e^{i\Vect{p}^\intercal \Vect{s}} \,
    D
    \left(
        \Vect{q} 
        - \frac{\Vect{s}}{2}
        , 
        \Vect{q} 
        + \frac{\Vect{s}}{2}
    \right)
    ,
    \label{eq:Wigner}
\end{equation}

\noindent where the integral is taken over $\Vect{q}$-space and $\doteq$ denotes definitions. (Note that vectors are interpreted as row vectors unless explicitly transposed via $^\intercal$, so $\Vect{p}^\intercal \Vect{s} = \Vect{p} \cdot \Vect{s} $.) The function $\Symb{D}(\Vect{q}, \Vect{p})$ is known as the Weyl symbol of $D(\Vect{q}, \Vect{q}')$. We shall now analyze \Eq{eq:WAVEintegral} in the short-wavelength limit using both traditional GO and MGO. Specifically, we assume in the following that the medium parameters, the wave envelope, and the wavelength itself vary slowly over a characteristic wavelength~\cite{Kravtsov90,Tracy14,Dodin19}.


\subsection{Traditional geometrical optics}

In traditional GO, $\psi(\Vect{q})$ is assumed to take the eikonal form given by
\begin{equation}
    \psi(\Vect{q}) = 
    \phi(\Vect{q}) \,
    e^{i \theta(\Vect{q}) } ,
    \label{eq:GOwave}
\end{equation}

\noindent where $\phi(\Vect{q})$ is a slowly varying envelope and $\theta(\Vect{q})$ is a rapidly varying phase. Then, to the lowest order, \Eqs{eq:WAVEintegral}-\eq{eq:GOwave} yield the local dispersion relation~\cite{Tracy14,Dodin19}
\begin{equation}
    \Symb{D}
    \left[
        \Vect{q}, \Vect{k}(\Vect{q})
    \right]
    = 0 ,
    \label{eq:GO1}
\end{equation}

\noindent along with the envelope transport equation
\begin{equation}
    \Vect{v}(\Vect{q})^\intercal \pd{\Vect{q}} \log  \phi(\Vect{q})
    = - \frac{1}{2} \pd{\Vect{q}} \cdot \Vect{v}(\Vect{q}) 
    .
    \label{eq:GO2}
\end{equation}

\noindent Here, the local wavevector $\Vect{k}(\Vect{q})$ and the local group velocity $\Vect{v}(\Vect{q})$ are defined as
\begin{equation}
    \Vect{k}(\Vect{q}) 
    \doteq
    \pd{\Vect{q}} \theta(\Vect{q})
    ,
    \quad
    \Vect{v}(\Vect{q})
    \doteq
    \left.
        \pd{\Vect{p}}
        \Symb{D}(\Vect{q}, \Vect{p})
    \right|_{\Vect{p} = \Vect{k}(\Vect{q})}
    .
\end{equation}

\noindent Hence, $\Vect{k}(\Vect{q})$ is irrotational, meaning that
\begin{equation}
    \pd{q_\ell} k_m = \pd{q_m} k_\ell 
    ,
    \quad 
    \ell,m = 1, \ldots, N .
\end{equation}

Equations \eq{eq:GO1} and \eq{eq:GO2} are commonly solved along the characteristic rays that satisfy
\begin{subequations}
    \label{eq:rays}
    \begin{align}
        \pd{\tau_1} \Vect{q}(\Vect{\tau})
        &= \pd{\Vect{k}} \Symb{D}
        \left[
            \Vect{q}(\Vect{\tau}),
            \Vect{k}(\Vect{\tau})
        \right]
        ,
        \\
        \pd{\tau_1} \Vect{k}(\Vect{\tau})
        &= - \pd{\Vect{q}} \Symb{D}
        \left[
            \Vect{q}(\Vect{\tau}),
            \Vect{k}(\Vect{\tau})
        \right] .
    \end{align}
\end{subequations}

\noindent For integrable systems, these rays trace out an $N$-D surface in phase space called the dispersion manifold. We  then define $\Vect{\tau} \doteq (\tau_1, \ldots, \tau_N)$ as coordinates on the dispersion manifold, with $\tau_1$ serving as the longitudinal coordinate along a ray. Initial conditions provide $\Vect{q}(0,\Vect{\tau}_\perp)$ and $\Vect{k}(0,\Vect{\tau}_\perp)$, where $\Vect{\tau}_\perp \doteq (\tau_2, \ldots, \tau_N)$. For example, $\tau_1$ can be the time variable or one of the spatial coordinates; then $\Vect{\tau}_\perp$ are the remaining spatial coordinates.

The wavevector $\Vect{k}(\Vect{q})$ is determined from the rays as
\begin{equation}
    \Vect{k}(\Vect{q})
    =
    \Vect{k}\left[ \Vect{\tau}(\Vect{q}) \right] ,
    \label{eq:kFIELD}
\end{equation}

\noindent where $\Vect{\tau}(\Vect{q})$ is the function inverse to $\Vect{q}(\Vect{\tau})$. Although \Eqs{eq:rays} show that rays cannot cross in phase space, their projections onto $\Vect{q}$-space can. As a result, $\Vect{\tau}(\Vect{q})$ is generally multivalued, meaning $\Vect{k}(\Vect{q})$ is as well. The envelope is constructed along the rays as
\begin{equation}
    \phi
    \left[
        \Vect{q}(\Vect{\tau})
    \right]
    = \phi
    \left[
        \Vect{q}(0,\Vect{\tau}_\perp) 
    \right]
    \sqrt
    {
        \frac
        { j(0, \Vect{\tau}_\perp) }
        { j(\Vect{\tau}) }
    }
    \label{eq:GOenv}
    ,
\end{equation}

\noindent where $\phi[\Vect{q}(0,\Vect{\tau}_\perp)]$ is set by initial conditions, and 
\begin{equation}
    j(\Vect{\tau})
    \doteq
    \det \pd{\Vect{\tau}} \Vect{q}(\Vect{\tau}) .
    \label{eq:JACq}
\end{equation}

\noindent Then, the total wavefield is constructed as
\begin{equation}
    \psi(\Vect{q}) = \sum_{\Vect{t} \in \Vect{\tau}(\Vect{q})} 
    \phi
    \left[
        \Vect{q}(\Vect{t})
    \right]
    \, 
    \exp
    \left[
        i \int \dd \Vect{q}^\intercal \,
        \Vect{k}(\Vect{q})
    \right] ,
    \label{eq:GOpsi}
\end{equation}

\noindent where the summation is taken over all branches of $\Vect{\tau}(\Vect{q})$. 

Clearly, $\psi(\Vect{q})$ diverges where $j(\Vect{t}) = 0$. Such points are caustics, and by \Eq{eq:JACq}, they occur where the dispersion manifold has a singular projection onto $\Vect{q}$-space. To extend GO modeling to caustics and the neighboring regions, \Ref{Lopez20a} proposed MGO, which we now describe.

\subsection{Metaplectic geometrical optics}

MGO also uses the rays provided by \Eqs{eq:rays} to solve \Eq{eq:GO1}. However, in MGO, the phase space is not fixed but instead is continually rotated, $(\Vect{q}, \Vect{p}) \to (\Vect{Q}_\Vect{t}, \Vect{P}_\Vect{t})$, such that the local projection of the dispersion manifold onto $\Vect{Q}_\Vect{t}$-space is always well-behaved. Accordingly, the envelope equation \eq{eq:GO2} is replaced with a similar envelope equation in the rotated frame that has no caustics by construction. This is done as follows.

Let us assume that the dispersion manifold $(\Vect{q}, \Vect{p}) = (\Vect{q}(\Vect{\tau}), \Vect{k}(\Vect{\tau}))$ has been obtained by integrating \Eqs{eq:rays} and consider the tangent plane at $\Vect{\tau} = \Vect{t}$. We can rotate our original phase space to align $\Vect{q}$-space with the tangent plane at $\Vect{t}$ ($\Vect{Q}_\Vect{t}$-space) using the following linear transformation:
\begin{equation}
    \Vect{Q}_\Vect{t} 
    = 
    \Mat{A}_\Vect{t} \Vect{q} 
    + \Mat{B}_\Vect{t} \Vect{p}
    ,
    \quad
    \Vect{P}_\Vect{t} 
    = 
    \Mat{C}_\Vect{t} \Vect{q} 
    + \Mat{D}_\Vect{t} \Vect{p} 
    ,
    \label{eq:linTRANS}
\end{equation}

\noindent where $\Vect{P}_\Vect{t}$ is Fourier-dual to $\Vect{Q}_\Vect{t}$ and the matrices $\Mat{A}_\Vect{t}$, $\Mat{B}_\Vect{t}$, $\Mat{C}_\Vect{t}$, and $\Mat{D}_\Vect{t}$ are all $N \times N$. In \Eq{eq:linTRANS}, we require that
\begin{equation}
    \Mat{S}_\Vect{t} 
    \doteq
    \begin{pmatrix}
        \Mat{A}_\Vect{t} & \Mat{B}_\Vect{t} \\
        \Mat{C}_\Vect{t} & \Mat{D}_\Vect{t}
    \end{pmatrix}
    \label{eq:sympMAT}
\end{equation}

\noindent be symplectic, that is,
\begin{equation}
    \Mat{S}_\Vect{t}
    \Mat{J} \,
    \Mat{S}_\Vect{t}^\intercal
    =
    \Mat{J}
    ,
    \quad
    \Mat{J} \doteq
    \begin{pmatrix}
        \OMat{N} & \IMat{N} \\
        - \IMat{N} & \OMat{N}
    \end{pmatrix} ,
    \label{eq:symplec}
\end{equation}

\noindent where $\OMat{N}$ and $\IMat{N}$ are respectively the $N \times N$ null and identity matrices. (A practical algorithm for computing $\Mat{S}_\Vect{t}$ from the ray trajectories using Gram--Schmidt orthogonalization is provided in \Ref{Lopez20a}.)

The transformation of the wavefield corresponding to the symplectic transformation \eq{eq:sympMAT} of the ray phase space is the MT~\cite{Littlejohn86a,Lopez19a}, sometimes called the linear canonical transform. The MT is a linear integral transformation from $\psi$ to a new function $\Psi$ given explicitly as
\begin{equation}
    \Psi(\Vect{Q})
    =
    \int \dd \Vect{q} \,
    M(\Vect{Q}, \Vect{q}) \,
    \psi(\Vect{q})
    ,
\end{equation}

\noindent where the MT kernel is given as
\begin{equation}
    M(\Vect{Q}, \Vect{q}) 
    \doteq
    \frac
    {
        \sigma
        \exp
        \left[ 
            i G(\Vect{q}, \Vect{Q})
        \right]
    }
    {
        (2\pi i)^{N/2}
        \sqrt{\det \Mat{B}}
    }
    ,
    \label{eq:METkern}
\end{equation}

\noindent and $G(\Vect{q}, \Vect{Q})$ is the following quadratic phase function:
\begin{align}
    G(\Vect{q}, \Vect{Q})
    &\doteq
    \frac{1}{2} \Vect{Q}^\intercal \Mat{D} \Mat{B}^{-1} \Vect{Q}
    - \Vect{Q}^\intercal \Mat{B}^{-\intercal} \Vect{q}
    + \frac{1}{2} \Vect{q}^\intercal \Mat{B}^{-1} \Mat{A} \Vect{q}
    .
    \label{eq:Geq}
\end{align}

\noindent (Here, $^{-\intercal}$ denotes the matrix inverse transpose.) Each $\Mat{S}$ actually has two corresponding MTs that differ by an overall sign, which is designated by $\sigma \doteq \pm 1$. Note that \Eq{eq:METkern} requires $\det \Mat{B} \neq 0$; this requirement will be lifted in \Sec{sec:singMET}. Also note that the MT reduces to the familiar FT when $\Mat{S} = \Mat{J}$.

Let $\Psi_\Vect{t}(\Vect{Q}_\Vect{t})$ be the MT of $\psi(\Vect{q})$ corresponding to $\Mat{S}_\Vect{t}$ of \Eq{eq:sympMAT} (for a chosen MT sign convention). 
In MGO, we assume that $\Psi_\Vect{t}(\Vect{Q}_\Vect{t})$ [not $\psi(\Vect{q})$] has the eikonal form
\begin{equation}
    \Psi_\Vect{t}(\Vect{Q}_\Vect{t}) 
    = \Phi_\Vect{t}(\Vect{Q}_\Vect{t}) \,
    e^{i \Theta_\Vect{t}(\Vect{Q}_\Vect{t})}
    ,
    \label{eq:MGOwave}
\end{equation}

\noindent where $\Phi_\Vect{t}$ is a slowly varying complex envelope and $\Theta_\Vect{t}$ is a phase that varies rapidly with the new coordinate $\Vect{Q}_\Vect{t}$; hence, $\Vect{K}_\Vect{t}(\Vect{Q}_\Vect{t}) \doteq \pd{\Vect{Q}_\Vect{t}} \Theta_\Vect{t}(\Vect{Q}_\Vect{t})$ is understood as the local wavevector in $\Vect{Q}_\Vect{t}$-space. The eikonal approximation \eq{eq:MGOwave} is facilitated by the fact that $\pd{\tau_1}\Vect{K}_{\Vect{t}} = - \pd{\Vect{Q}_\Vect{t}} \fourier{\Symb{D}} = \Vect{0}$ at the tangent point by definition, where
\begin{equation}
    \fourier{\Symb{D}}
    (\Vect{Q}_\Vect{t}, \Vect{P}_\Vect{t})
    =
    \Symb{D}
    (
        \Mat{D}_\Vect{t}^\intercal \Vect{Q}_\Vect{t}
        - \Mat{B}_\Vect{t}^\intercal \Vect{P}_\Vect{t}
        ,
        - \Mat{C}_\Vect{t}^\intercal \Vect{Q}_\Vect{t}
        + \Mat{A}_\Vect{t}^\intercal \Vect{P}_\Vect{t}
    )
    ,
\end{equation}

\noindent is the Weyl symbol of the dispersion operator in the new coordinates. (We also assume that $\pd{\tau_1}\Vect{K}_{\Vect{t}}$ is slowly varying in the neighborhood of the tangent point.)

As shown in \Ref{Lopez20a}, taking the MT of \Eq{eq:WAVEintegral} and performing the standard GO procedure in $\Vect{Q}_\Vect{t}$-space using \Eq{eq:MGOwave} yields equations similar to \Eqs{eq:GO1} and \eq{eq:GO2} in the new variables. Specifically, the local dispersion relation has the form
\begin{equation}
    \fourier{\Symb{D}}
    \left[
        \Vect{Q}_\Vect{t}, 
        \Vect{K}_\Vect{t}(\Vect{Q}_\Vect{t})
    \right]
    = 0 ,
    \label{eq:MGO1}
\end{equation}

\noindent and the envelope transport equation becomes
\begin{equation}
    \Vect{V}_\Vect{t}(\Vect{Q}_\Vect{t})^\intercal \pd{\Vect{Q}_\Vect{t}} \log  \Phi_\Vect{t}(\Vect{Q}_\Vect{t})
    = - \frac{1}{2} \pd{\Vect{Q}_\Vect{t}} \cdot \Vect{V}_\Vect{t}(\Vect{Q}_\Vect{t}) .
    \label{eq:MGO2}
\end{equation}

\noindent Here, we have introduced
\begin{equation}
    \Vect{V}_\Vect{t}(\Vect{Q}_\Vect{t})
    \doteq
    \left.
    \pd{\Vect{P}_\Vect{t}}
        \fourier{\Symb{D}}(\Vect{Q}_\Vect{t},
        \Vect{P}_\Vect{t})
    \right|_
    {
        \Vect{P}_\Vect{t} 
        = \Vect{K}_\Vect{t}(\Vect{Q}_\Vect{t})
    } ,
\end{equation}

\noindent which is the corresponding group velocity.

If \Eq{eq:GO1} is satisfied along the original rays, then \Eq{eq:MGO1} is satisfied along the rotated rays given by
\begin{subequations}
    \label{eq:METrays}
    \begin{align}
        \Vect{Q}_\Vect{t}(\Vect{\tau})
        &=
        \Mat{A}_\Vect{t} \Vect{q}(\Vect{\tau})
        + \Mat{B}_\Vect{t} \Vect{k}(\Vect{\tau})
        ,
        \\
        \Vect{K}_\Vect{t}(\Vect{\tau})
        &=
        \Mat{C}_\Vect{t} \Vect{q}(\Vect{\tau})
        + \Mat{D}_\Vect{t} \Vect{k}(\Vect{\tau}) .
    \end{align}
\end{subequations}

\noindent Hence, the dispersion manifold simply rotates with the ambient phase space%
\footnote{This is analogous to Wigner functions being simply rotated by fractional FTs~\cite{Lohmann93}.}.%

Before solving \Eq{eq:MGO2}, it is convenient to renormalize $\Psi_\Vect{t}(\Vect{Q}_\Vect{t})$ by its value at $\Vect{Q}_\Vect{t}(\Vect{t})$; that is, let
\begin{equation}
    \Psi_\Vect{t}(\Vect{Q}_\Vect{t}) 
    = \alpha_\Vect{t} 
    \Phi_\Vect{t}(\Vect{Q}_\Vect{t})
    e^{i \Theta_\Vect{t}(\Vect{Q}_\Vect{t})}
    ,
    \quad
    \alpha_\Vect{t}
    \doteq 
    \Psi_\Vect{t}
    \left[
        \Vect{Q}_\Vect{t}(\Vect{t})
    \right] ,
\end{equation}

\noindent and require
\begin{equation}
    \Phi_\Vect{t}
    \left[
        \Vect{Q}_\Vect{t}(\Vect{t})
    \right]
    = 1
    ,
    \quad
    \Theta_\Vect{t}
    \left[
        \Vect{Q}_\Vect{t}(\Vect{t})
    \right]
    = 0 .
    \label{eq:MGOconstr}
\end{equation}

\noindent Then, analogous to \Eq{eq:GOenv}, \Eq{eq:MGO2} is solved to yield
\begin{equation}
    \Phi_\Vect{t}
    \left[
        \Vect{Q}_\Vect{t}(\Vect{\tau})
    \right]
    = 
    \Phi_\Vect{t}
    \left[
        \Vect{Q}_\Vect{t}(t_1, \Vect{\tau}_\perp)
    \right]
    \sqrt
    {
        \frac
        {
            J_\Vect{t}(t_1, \Vect{\tau}_\perp)
        }
        {
            J_\Vect{t}(\Vect{\tau})
        }
    }
    ,
    \label{eq:MGOphi}
\end{equation}

\noindent where $\Phi_\Vect{t}[\Vect{Q}_\Vect{t}(t_1, \Vect{\tau}_\perp)]$ is set by initial conditions subject to \Eq{eq:MGOconstr}, and
\begin{equation}
    J_\Vect{t}(\Vect{\tau})
    \doteq
    \det \pd{\Vect{\tau}} \Vect{Q}_\Vect{t}(\Vect{\tau})
    .
    \label{eq:MGOjac}
\end{equation}

\noindent The phase can also be immediately determined as
\begin{equation}
    \Theta_\Vect{t}(\Vect{Q}_\Vect{t})
    =
    \int_{\Vect{Q}_\Vect{t}(\Vect{t})}^{\Vect{Q}_\Vect{t}}
    \dd \Vect{Q}^\intercal \Vect{K}_\Vect{t}(\Vect{Q})
    ,
    \label{eq:MGOphase}
\end{equation}

\noindent where the line integral is taken over any path with the specified endpoints, and $\Vect{K}_\Vect{t}(\Vect{Q}_\Vect{t})$ is constructed as
\begin{equation}
    \Vect{K}_\Vect{t}(\Vect{Q}_\Vect{t}) 
    = 
    \Vect{K}_\Vect{t}
    \left[ 
        \Vect{\tau}(\Vect{Q}_\Vect{t})
    \right] ,
    \label{eq:Kdef}
\end{equation}

\noindent where $\Vect{\tau}(\Vect{Q}_\Vect{t})$ is the function inverse to $\Vect{Q}_\Vect{t}(\Vect{\tau})$. Note that $\Vect{K}_\Vect{t}(\Vect{Q}_\Vect{t})$ will generally be multivalued, so it must be restricted to the branch satisfying $\Vect{K}_\Vect{t}[\Vect{Q}_\Vect{t}(\Vect{t})] = \Vect{K}_\Vect{t}(\Vect{t})$.

By continuity, $\alpha_\Vect{t}$ evolves along the rays as
\begin{equation}
    \alpha_\Vect{t} 
    = 
    \alpha_{\left(0,\Vect{t}_\perp \right)} 
    \exp
    \left[
        \int_0^{t_1} \dd h \, \eta_{\left(h, \Vect{t}_\perp \right)}
    \right] ,
    \label{eq:MGOalpha}
\end{equation}

\noindent where $\alpha_{\left(0,\Vect{t}_\perp \right)}$ is determined by initial conditions and
\begin{align}
    \eta_\Vect{t}
    &\doteq
    \frac{i}{2} \Vect{K}^\intercal _\Vect{t}(\Vect{t}) 
    \Mat{W}_\Vect{t} \Vect{K}_\Vect{t}(\Vect{t})
    - \frac{i}{2} \Vect{Q}^\intercal _\Vect{t}(\Vect{t}) 
    \Mat{U}_\Vect{t} \Vect{Q}_\Vect{t}(\Vect{t})
    - \frac{1}{2}
    \Tr\left(\Mat{V}_\Vect{t} \right) 
    \nonumber\\
    &\hspace{4mm}+\left[
        \pd{h} \Vect{Q}_\Vect{t}(\Vect{t}) 
        - \Mat{V}_\Vect{t}^\intercal \Vect{Q}_\Vect{t}(\Vect{t}) 
        - \Mat{W}_\Vect{t}^\intercal \Vect{K}_\Vect{t}(\Vect{t})
        \nullFrac
    \right]^\intercal 
    \nonumber\\
    &\hspace{8mm}\times
    \left\{
        \pd{\Vect{Q}} \Phi_\Vect{t}\left[
            \Vect{Q}_\Vect{t}(\Vect{t})
        \right]
        + i \Vect{K}_\Vect{t}(\Vect{t}) 
        \nullFrac
    \right\}
    .
    \label{eq:MGOeta}
\end{align}

\noindent Here, the $N \times N$ matrices $\Mat{U}_\Vect{t}$, $\Mat{V}_\Vect{t}$, and $\Mat{W}_\Vect{t}$ are defined via
\begin{equation}
    \left(\pd{h} \Mat{S}_\Vect{t}\right)\Mat{S}_\Vect{t}^{-1}
    \doteq
    \begin{pmatrix}
        \Mat{V}_\Vect{t}^\intercal & \Mat{W}_\Vect{t} \\
        - \Mat{U}_\Vect{t} & - \Mat{V}_\Vect{t}
    \end{pmatrix}
    ,
    \label{eq:MGOuvw}
\end{equation}

\noindent where the directional derivative, defined as $h \, \pd{h} \doteq \Vect{h}^\intercal \pd{\Vect{t}}$, should be interpreted as a total derivative acting on arguments and subscripts containing $\Vect{t}$. (An alternate algorithm for evolving $\alpha_\Vect{t}$ based on successive applications of a near-identity MT~\cite{Lopez19a} is provided in \Ref{Lopez20a}.) 

Applying an inverse MT then maps $\Psi_\Vect{t}(\Vect{Q}_\Vect{t})$ to a function on $\Vect{q}$-space, denoted $\psi_\Vect{t}(\Vect{q})$, as
\begin{equation}
    \psi_\Vect{t}(\Vect{q})
    =
    \int \dd \Vect{Q}_\Vect{t} \, 
    M_\Vect{t}^{-1}(\Vect{q}, \Vect{Q}_\Vect{t}) \,
    \Psi_\Vect{t}(\Vect{Q}_\Vect{t})
    ,
    \label{eq:metINV}
\end{equation}

\noindent where the inverse MT kernel is given as
\begin{equation}
    M_\Vect{t}^{-1}(\Vect{q}, \Vect{Q}_\Vect{t}) 
    \doteq
    \frac
    {
        \sigma_\Vect{t}
        \exp
        \left[ 
            - i G_\Vect{t}(\Vect{q}, \Vect{Q}_\Vect{t})
        \right]
    }
    {
        (-2\pi i)^{N/2}
        \sqrt{\det \Mat{B}_\Vect{t}}
    }
    .
    \label{eq:invMETkern}
\end{equation}

\noindent The overall sign factor $\sigma_\Vect{t}$ must change whenever $\det \Mat{B}_\Vect{t}$ crosses the branch cut of the square root to maintain continuity. This change in $\sigma_\Vect{t}$ is related to the discrete phase jumps a wavefield experiences upon traversing a caustic.

At this point, semiclassical methods like GO and Maslov's method traditionally evaluate \Eq{eq:metINV} using the stationary-phase approximation (SPA) about the ray contribution $\Vect{Q}_\Vect{t} = \Vect{Q}_\Vect{t}(\Vect{t})$~\cite{Heller77a}. However, the SPA fails when saddlepoints are close together~\cite{Chester57}, as occurs near caustics. To remedy this, note that under fairly general conditions, integrals like \Eq{eq:metINV} can be evaluated on the union of steepest-descent contours through some subset of saddlepoints in complex $\Vect{Q}_\Vect{t}$-space~\cite{Bleistein86}. By integrating \Eq{eq:metINV} only along the steepest-descent contour through $\Vect{Q}_\Vect{t} = \Vect{Q}_\Vect{t}(\Vect{t})$ rather than the entire set, we can isolate the desired ray contribution in a manner that is asymptotically equivalent to the SPA but is also well-behaved at caustics. (In this regard, we can also define the `saddlepoint contribution' to an integral as the result of integrating along the corresponding steepest-descent contour.) Hence, the variable shift $\Vect{\epsilon} \doteq \Vect{Q}_\Vect{t} - \Vect{Q}_\Vect{t}(\Vect{t})$ yields
\begin{equation}
    \psi_\Vect{t}(\Vect{q})
    =
    \frac{
        \sigma_\Vect{t} \, \alpha_\Vect{t} 
        \exp
        \left[ 
            - \frac{i}{2} \beta_\Vect{t}(\Vect{q}) 
        \right]
    }
    {
        (-2\pi i)^{N/2} \sqrt{ \det{\Mat{B}_\Vect{t}}}
    } \,
    \Upsilon_\Vect{t}(\Vect{q})
    ,
    \label{eq:MGOpsiT}
\end{equation}

\noindent where we have defined
\begin{subequations}
    \begin{align}
        \beta_\Vect{t}(\Vect{q})
        &\doteq 2 \, G_\Vect{t}[\Vect{q}, \Vect{Q}_\Vect{t}(\Vect{t})]
        , \\
        \Upsilon_\Vect{t}(\Vect{q})
        &\doteq 
            \int_{\cont{0}} \dd \Vect{\epsilon} \,
            \Psi_\Vect{t}\left[ 
                \Vect{\epsilon} + \Vect{Q}_\Vect{t}(\Vect{t})
            \right] 
            \exp\left[ 
                - i \gamma_\Vect{t}(\Vect{\epsilon}, \Vect{q})
            \right] 
        , \\
        \gamma_\Vect{t}(\Vect{\epsilon}, \Vect{q}) 
        &\doteq 
            \frac{1}{2} \Vect{\epsilon}^\intercal \Mat{D}_\Vect{t} \Mat{B}_\Vect{t}^{-1} \Vect{\epsilon}
            +
            \Vect{\epsilon}^\intercal
            \Mat{B}_\Vect{t}^{-\intercal}
            \left[   
                \Mat{D}_\Vect{t}^\intercal \Vect{Q}_\Vect{t}(\Vect{t})
                -
                \Vect{q}
            \right]
        ,
    \end{align}
\end{subequations}

\noindent and $\cont{0}$ is the steepest-descent contour through $\Vect{\epsilon} = \Vect{0}$. Finally, $\psi(\Vect{q})$ is reconstructed by summing over all branches of the dispersion manifold:
\begin{equation}
    \psi(\Vect{q}) 
    = 
    \sum_{\Vect{t} \in \Vect{\tau}(\Vect{q})}
    \psi_\Vect{t}
    \left[ \Vect{q}(\Vect{t}) \right] .
    \label{eq:MGOpsi}
\end{equation}

Equation \eq{eq:MGOpsi} can accurately model the wavefield incident on an isolated fold caustic or bounded between a pair of fold caustics in 1-D. However, \Eq{eq:MGOpsi} cannot model a wavefield whose dispersion manifold has $\det{\Mat{B}_\Vect{t}} = 0$ over a finite domain (examples of which are discussed in \Sec{sec:ex}). We call such ray patterns `quasiuniform'. To enable MGO to model quasiuniform ray patterns, the restriction that $\det \Mat{B}_\Vect{t} \neq 0$ must be lifted. The corresponding theory is discussed in the next section.

\section{Metaplectic geometrical optics for quasiuniform ray patterns}
\label{sec:MGO2}

\subsection{Singular metaplectic transforms}
\label{sec:singMET}

The MT kernel $M(\Vect{Q}, \Vect{q}; \Mat{S})$ that corresponds to a symplectic matrix $\Mat{S}$ with $\det \Mat{B} = 0$ can be considered as a limiting case~\cite{Littlejohn86a,Lopez19a}
\begin{equation}
    M(\Vect{Q}, \Vect{q}; \Mat{S}) = \lim_{\varepsilon \to 0} M(\Vect{Q}, \Vect{q}; \Mat{S}_\varepsilon) ,
    \label{eq:MTlim}
\end{equation}

\noindent where $\Mat{S}_\varepsilon$ is a symplectic matrix that has $\Mat{S}$ as a limit at $\varepsilon \to 0$. (Here and further, we omit the subscript $_\Vect{t}$ for ease of notation.) For example, we can adopt
\begin{equation}
    \Mat{S}_\varepsilon 
    = 
    \begin{pmatrix}
        \Mat{A} & \Mat{B} + \varepsilon \Mat{A} \\
        \Mat{C} & \Mat{D} + \varepsilon \Mat{C}
    \end{pmatrix} ,
\end{equation}

\noindent whose symplecticity (to all orders in $\varepsilon$) can be readily verified by definition \eq{eq:symplec}.

To show that $\det \left(\Mat{B} + \varepsilon \Mat{A} \right) \neq 0$ and subsequently compute the limit in \Eq{eq:MTlim}, it is useful to perform a singular value decomposition (SVD) of $\Mat{B}$. Let $\rho$ and $\varsigma \doteq N - \rho$ be the rank and corank of $\Mat{B}$ respectively. Then, the SVD of $\Mat{B}$ takes the form
\begin{equation}
    \Mat{B} = \Mat{L} \, \widetilde{\Mat{B}} \, \Mat{R}^\intercal ,
\end{equation}

\noindent where $\widetilde{\Mat{B}}$ is a diagonal matrix given by
\begin{equation}
    \widetilde{\Mat{B}}
    =
    \begin{pmatrix}
        \Mat{\Lambda}_{\rho \rho} & \OMat{ \rho \varsigma} \\
        \OMat{\varsigma \rho} & \OMat{\varsigma \varsigma}
    \end{pmatrix}
\end{equation}

\noindent (the submatrices with subscript $_{mn}$ are size $m \times n$) and $\Mat{\Lambda}_{\rho \rho}$ is a diagonal matrix that has all nonzero singular values of $\Mat{B}$ on its diagonal:
\begin{equation}
    \Mat{\Lambda}_{\rho \rho}
    \doteq
    \begin{pmatrix}
        \lambda_1 & & \\
        & \ddots & \\
        & & \lambda_\rho
    \end{pmatrix} .
\end{equation}

\noindent Note that $\det \Mat{\Lambda}_{\rho \rho} \neq 0$ by definition. The matrices $\Mat{L}$ and $\Mat{R}$ are both orthogonal and can be written as
\begin{equation}
    \Mat{L} 
    = 
    \begin{pmatrix}
        \uparrow &  & \uparrow \\[1mm]
        \unit{\Vect{\ell}}_1 & \ldots & \unit{\Vect{\ell}}_N \\[1mm]
        \downarrow &  & \downarrow
    \end{pmatrix}
    ,
    \quad
    \Mat{R} 
    = 
    \begin{pmatrix}
        \uparrow &  & \uparrow \\[1mm]
        \unit{\Vect{r}}_1 & \ldots & \unit{\Vect{r}}_N \\[1mm]
        \downarrow &  & \downarrow
    \end{pmatrix}
    .
\end{equation}

\noindent The columns of these matrices are, respectively, the left singular vectors $\{ \unit{\Vect{\ell}}_j \}$ and right singular vectors $\{ \unit{\Vect{r}}_j \}$ of $\Mat{B}$, which are mutually orthonormal:
\begin{equation}
    \unit{\Vect{\ell}}_j^\intercal \unit{\Vect{\ell}}_k = \delta_{jk}
    ,
    \quad
    \unit{\Vect{r}}_j^\intercal \unit{\Vect{r}}_k = \delta_{jk} 
    .
\end{equation}

Let us similarly define
\begin{equation}
    \widetilde{\Mat{A}} \doteq \Mat{L}^\intercal \Mat{A} \Mat{R}
    ,
    \quad
    \widetilde{\Mat{C}} \doteq \Mat{L}^\intercal \Mat{C} \Mat{R}
    ,
    \quad
    \widetilde{\Mat{D}} \doteq \Mat{L}^\intercal \Mat{D} \Mat{R} 
    .
\end{equation}

\noindent As shown in \App{app:Matrices}, these matrices have the form
\begin{equation}
    \widetilde{\Mat{A}} 
    =
    \begin{pmatrix}
        \Mat{a}_{\rho \rho} & \Mat{a}_{\rho \varsigma} \\
        \OMat{\varsigma \rho} & \Mat{a}_{\varsigma \varsigma}
    \end{pmatrix}
    ,
    \quad
    \widetilde{\Mat{C}} 
    =
    \begin{pmatrix}
        \Mat{c}_{\rho \rho} & \Mat{c}_{\rho \varsigma} \\
        \Mat{c}_{\varsigma \rho} & \Mat{c}_{\varsigma \varsigma}
    \end{pmatrix}
    , 
    \quad
    \widetilde{\Mat{D}} 
    = 
    \begin{pmatrix}
        \Mat{d}_{\rho \rho} & \OMat{\rho \varsigma} \\
        \Mat{d}_{\varsigma \rho} & \Mat{a}_{\varsigma \varsigma}^{-\intercal}
    \end{pmatrix}
    .
\end{equation}

\noindent Note that $\Mat{a}_{\varsigma \varsigma}$ is invertible. Hence, we compute
\begin{align}
    \det
    \left(
        \Mat{B} + \varepsilon \Mat{A}
    \right)
    &=
    \det
    \begin{pmatrix}
        \Mat{\Lambda}_{\rho \rho} + \varepsilon \Mat{a}_{\rho \rho} 
        & 
        \varepsilon \Mat{a}_{\rho \varsigma} 
        \\
        \OMat{\varsigma \rho} 
        & 
        \varepsilon \Mat{a}_{\varsigma \varsigma}
    \end{pmatrix}
    \nonumber\\
    &=
    \det 
    \left( 
        \Mat{\Lambda}_{\rho \rho} + \varepsilon \Mat{a}_{\rho \rho} 
    \right)
    \det
    \left( 
        \varepsilon \Mat{a}_{\varsigma \varsigma}
    \right)
    \nonumber\\
    &\approx
    \varepsilon^\varsigma
    \det \Mat{\Lambda}_{\rho \rho} \,
    \det \Mat{a}_{\varsigma \varsigma}
    ,
    \label{eq:detBmod}
\end{align}

\noindent where we have used $\det \Mat{L} = \det \Mat{R} = 1$. Since $\det \Mat{a}_{\varsigma \varsigma} \ne 0$ and $\det \Mat{\Lambda}_{\rho \rho} \ne 0$ by definition, $\det(\Mat{B} + \varepsilon \Mat{A})$ is nonzero for finite $\varepsilon$. (We adopt the convention that $0 \times 0$ matrices have unit determinant.) 

By \Eqs{eq:METkern} and \eq{eq:detBmod}, we obtain to leading order in $\varepsilon$
\begin{align}
    M(\Vect{Q}, \Vect{q}; \Mat{S}_\varepsilon)
    &\approx 
    \frac
    {
        \sigma \, 
        \varepsilon^{-\varsigma/2} 
        \exp
        \left[
            i \, g(\Vect{q}_\rho, \Vect{Q})
        \right]
    }
    {
        (2 \pi i)^{N/2}
        \sqrt
        {
            \det \Mat{\Lambda}_{\rho \rho}
            \det \Mat{a}_{\varsigma \varsigma}
        }
    }
    \nonumber\\
    &\hspace{4mm}\times
    \exp
    \left(
        \frac{i}{2\varepsilon}
        \left|
            \Vect{q}_\varsigma - \Mat{a}_{\varsigma \varsigma}^{-1} \Vect{Q}_\varsigma
        \right|^2
    \right)
    ,
    \label{eq:MTepsilon}
\end{align}

\noindent where we have defined 
\begin{align}
    g(\Vect{q}_\rho, \Vect{Q})
    \doteq
    \frac{1}{2} \Vect{q}_\rho^\intercal \, \Mat{\Lambda}_{\rho \rho}^{-1} \Mat{a}_{\rho \rho} \, \Vect{q}_\rho
    &- \Vect{q}_\rho^\intercal \, \Mat{M}_1 \, \Mat{L}^\intercal \Vect{Q}
    \nonumber\\
    &+ \frac{1}{2} \Vect{Q}^\intercal \, \Mat{L} \Mat{M}_2 \Mat{L}^\intercal \, \Vect{Q}
    ,
\end{align}

\noindent along with the matrices 
\begin{subequations}
    \label{eq:M12}
    \begin{align}
        \Mat{M}_1
        &\doteq
        \begin{pmatrix}
            \Mat{\Lambda}_{\rho \rho}^{-1} 
            & 
            -\Mat{\Lambda}_{\rho \rho}^{-1} \Mat{a}_{\rho \varsigma} \Mat{a}_{\varsigma \varsigma}^{-1} 
        \end{pmatrix}
        ,
        \\
        \Mat{M}_2
        &\doteq
        \begin{pmatrix}
            \Mat{d}_{\rho \rho} \, \Mat{\Lambda}_{\rho \rho}^{-1}
            &
            \Mat{\Lambda}_{\rho \rho}^{-1} \Mat{d}_{\varsigma \rho}^\intercal
            \\[1mm]
            \Mat{d}_{\varsigma \rho} \Mat{\Lambda}_{\rho \rho}^{-1}
            &
            \Mat{c}_{\varsigma \varsigma} \Mat{a}_{\varsigma \varsigma}^{-1} - \Mat{d}_{\varsigma \rho} \Mat{\Lambda}_{\rho \rho}^{-1} \Mat{a}_{\rho \varsigma} \Mat{a}_{\varsigma \varsigma}^{-1}
        \end{pmatrix}
    \end{align}
\end{subequations}

\noindent and the vector decompositions
\begin{equation}
    \Mat{R}^\intercal\Vect{q}
    =
    \begin{pmatrix}
        \Vect{q}_\rho \\[1mm]
        \Vect{q}_\varsigma
    \end{pmatrix}
    , \quad
    \Mat{L}^\intercal \Vect{Q}
    =
    \begin{pmatrix}
        \Vect{Q}_\rho \\[1mm]
        \Vect{Q}_\varsigma
    \end{pmatrix}
    .
    \label{eq:Qproj}
\end{equation}

\noindent Note that $\Mat{M}_2$ and $\Mat{\Lambda}_{\rho \rho}^{-1} \Mat{a}_{\rho \rho}$ are symmetric [\Eqs{eq:aSUBsymm}, \eq{eq:dSUBsymm}, and \eq{eq:cSUBsymm}]. Also note that $\Mat{M}_1$ is size $\rho \times N$, $\Mat{M}_2$ is size $N \times N$, and any vector $\Vect{\nu}_m$ is size $m \times 1$. [\App{app:MTphase} provides details for the derivation of \Eq{eq:MTepsilon}.] Then, using \Eq{eq:MTlim} along with
\begin{align}
    &\lim_{\varepsilon \to 0} 
    \varepsilon^{-\varsigma/2}
    \exp
    \left(
        \frac{i}{2 \varepsilon}
        \left| 
            \Vect{q}_\varsigma
            - \Mat{a}_{\varsigma \varsigma}^{-1} \Vect{Q}_\varsigma
        \right|^2
    \right)
    \nonumber\\
    &\hspace{32mm}=
    (2\pi i)^{\varsigma/2} \,
    \delta
    \left(
        \Vect{q}_\varsigma
        - \Mat{a}_{\varsigma \varsigma}^{-1} \Vect{Q}_\varsigma
    \right) ,
\end{align}

\noindent we obtain the limit of the MT kernel at $\det \Mat{B} \to 0$:
\begin{align}
    M(\Vect{Q}, \Vect{q})
    &= 
    \frac
    {
        \sigma \, 
        \exp
        \left[
            i \, g(\Vect{q}_\rho, \Vect{Q})
        \right]
        \,
        \delta 
        \left( 
            \Vect{q}_\varsigma - \Mat{a}_{\varsigma \varsigma}^{-1} \Vect{Q}_\varsigma
        \right)
    }
    {
        (2 \pi i)^{\rho/2} 
        \sqrt
        {
            \det \Mat{\Lambda}_{\rho \rho}
            \det \Mat{a}_{\varsigma \varsigma}
        }
    }
    ,
    \label{eq:METkernB}
\end{align}

\noindent where, for brevity, we no longer mention the dependence of $M$ on the symplectic matrix explicitly.

Following straightforward delta-function manipulations, we obtain the inverse MT kernel when $\det \Mat{B} = 0$:
\begin{equation}
    M^{-1}(\Vect{q}, \Vect{Q})
    = 
    \frac
    {
        \sigma \, 
        \exp
        \left[
            - i \,
            \widetilde{g}(\Vect{Q}_\rho, \Vect{q})
        \right]
        \, 
        \delta
        \left(
            \Vect{Q}_\varsigma - \Mat{a}_{\varsigma \varsigma} \Vect{q}_\varsigma
        \right)
    }
    {
        (-2\pi i)^{\rho/2}
        \sqrt
        {
            \det \Mat{\Lambda}_{\rho \rho}
            \det \Mat{a}_{\varsigma \varsigma}^{-1}
        }
    }
    ,
    \label{eq:invMETkernB}
\end{equation}

\noindent where we have defined 
\begin{align}
    \widetilde{g}(\Vect{Q}_\rho, \Vect{q})
    \doteq
    \frac{1}{2} \Vect{Q}_\rho^\intercal \, \Mat{d}_{\rho \rho} \Mat{\Lambda}_{\rho \rho}^{-1} \, \Vect{Q}_\rho
    &- \Vect{Q}_\rho^\intercal \, \Mat{M}_3 \Mat{R}^\intercal \Vect{q}
    \nonumber\\
    &+ \frac{1}{2} \Vect{q}^\intercal \, \Mat{R} \Mat{M}_4 \Mat{R}^\intercal \, \Vect{q}
    ,
\end{align}

\noindent along with the matrices
\begin{subequations}
    \begin{align}
        \Mat{M}_3 
        &\doteq
        \begin{pmatrix}
            \Mat{\Lambda}_{\rho \rho}^{-1}
            &
            - \Mat{\Lambda}_{\rho \rho}^{-1} \Mat{d}_{\varsigma \rho}^\intercal \Mat{a}_{\varsigma \varsigma}
        \end{pmatrix}
        ,
        \\
        \Mat{M}_4
        &\doteq
        \begin{pmatrix}
            \Mat{\Lambda}_{\rho \rho}^{-1} \Mat{a}_{\rho \rho}
            & \Mat{\Lambda}_{\rho \rho}^{-1} \Mat{a}_{\rho \varsigma}
            \\[1mm]
            \Mat{a}_{\rho \varsigma}^\intercal \Mat{\Lambda}_{\rho \rho}^{-1}
            &
            \Mat{a}_{\varsigma \varsigma}^\intercal \Mat{c}_{\varsigma \varsigma} - \Mat{a}_{\varsigma \varsigma}^\intercal \Mat{d}_{\varsigma \rho} \Mat{\Lambda}_{\rho \rho}^{-1} \Mat{a}_{\rho \varsigma}
        \end{pmatrix}
        .
    \end{align}
\end{subequations}

\noindent Note that $\Mat{M}_4$ and $\Mat{d}_{\rho \rho} \Mat{\Lambda}_{\rho \rho}^{-1}$ are symmetric [\Eqs{eq:aSUBsymm}, \eq{eq:dSUBsymm}, and \eq{eq:cSUBsymm}]. Also, $\Mat{M}_3$ is size $\rho \times N$ while $\Mat{M}_4$ is size $N \times N$. Lastly, we choose the following branch-cut convention: $\text{arg}(i) = \pi/2$ and $\text{arg}(\det \Mat{\Lambda}_{\rho \rho} \det\Mat{a}_{\varsigma \varsigma}) \in (-\pi, \pi]$ in \Eq{eq:METkernB}; $\text{arg}(-i) = -\pi/2$ and $\text{arg}(\det \Mat{\Lambda}_{\rho \rho} \det\Mat{a}_{\varsigma \varsigma}^{-1}) \in [-\pi, \pi)$ in \Eq{eq:invMETkernB}.


\subsection{Singular metaplectic geometrical optics}
\label{sec:singMGO}

Let us now incorporate the general representation for the inverse MT given by \Eq{eq:invMETkernB} into the MGO formalism. We emphasize that \Eq{eq:invMETkernB} is valid for all values of $\det \Mat{B}$. Using \Eqs{eq:metINV}, \eq{eq:Qproj}, and \eq{eq:invMETkernB} yields
\begin{align}
    \psi_\Vect{t}(\Vect{q})
    &=
    \mc{N}_\Vect{t}(\Vect{q})
    \int 
    \dd \Vect{Q}_\rho \,
    \dd \Vect{Q}_\varsigma \,
    \delta
    \left(
        \Vect{Q}_\varsigma - \Mat{a}_{\varsigma \varsigma} \Vect{q}_\varsigma
    \right)
    \, \Psi_\Vect{t}
    \left[
        \Mat{L} 
        \begin{pmatrix}
            \Vect{Q}_\rho \\[1mm]
            \Vect{Q}_\varsigma
        \end{pmatrix} 
    \right]
    \nonumber\\
    &\times
    \exp
    \left( 
        - \frac{i}{2} \Vect{Q}_\rho^\intercal \, \Mat{d}_{\rho \rho} \Mat{\Lambda}_{\rho \rho}^{-1} \, \Vect{Q}_\rho
        + i \Vect{Q}_\rho^\intercal \, \Mat{M}_4 \Mat{R}^\intercal \, \Vect{q}
    \right)
    ,
\end{align}

\noindent where we have substituted $\Vect{Q}$ with $\Vect{Q}_\rho$ and $\Vect{Q}_\varsigma$ as
\begin{equation}
    \Vect{Q}
    = \Mat{L}
    \begin{pmatrix}
        \Vect{Q}_\rho \\[1mm]
        \Vect{Q}_\varsigma
    \end{pmatrix} 
    ,
    \quad 
    \dd \Vect{Q} 
    = \dd \left( \Mat{L} \Vect{Q} \right)
    = \dd \Vect{Q}_\rho \, \dd \Vect{Q}_\varsigma
\end{equation}

\noindent and defined the prefactor
\begin{equation}
    \mc{N}_\Vect{t}(\Vect{q}) 
    \doteq 
    \frac
    {
        \sigma_\Vect{t} \,
        \alpha_\Vect{t} \,
        \exp
        \left(
            - \frac{i}{2} \Vect{q}^\intercal \, \Mat{R} \Mat{M}_4 \Mat{R}^\intercal \, \Vect{q}
        \right)
    }
    {
        (- 2\pi i)^{\rho/2}
        \sqrt
        {
            \det \Mat{\Lambda}_{\rho \rho} \,
            \det \Mat{a}_{\varsigma \varsigma}^{-1}
        }
    }
    .
\end{equation}

\noindent (As a reminder, all matrices depend on $\Vect{t}$.) The integration over $\Vect{Q}_\varsigma$ is immediately performed to yield
\begin{align}
    \psi_\Vect{t}(\Vect{q})
    &=
    \mc{N}_\Vect{t}(\Vect{q}) 
    \int 
    \dd \Vect{Q}_\rho \,
    \Psi_\Vect{t}
    \left[
        \Mat{L} 
        \begin{pmatrix}
            \Vect{Q}_\rho \\[1mm]
            \Mat{a}_{\varsigma \varsigma} \Vect{q}_\varsigma
        \end{pmatrix} 
    \right]
    \nonumber\\
    &\times
    \exp
    \left( 
        - \frac{i}{2} \Vect{Q}_\rho^\intercal \, \Mat{d}_{\rho \rho} \Mat{\Lambda}_{\rho \rho}^{-1} \, \Vect{Q}_\rho
        + i \Vect{Q}_\rho^\intercal \, \Mat{M}_3 \Mat{R}^\intercal \, \Vect{q}
    \right)
    .
\end{align}

The phase of the integrand is stationary where
\begin{equation}
    \pd{\Vect{Q}_\rho} \Theta_\Vect{t}
    \left[
        \Mat{L} 
        \begin{pmatrix}
            \Vect{Q}_\rho \\[1mm]
            \Mat{a}_{\varsigma \varsigma} \Vect{q}_\varsigma
        \end{pmatrix} 
    \right]
    + \Mat{M}_3 \Mat{R}^\intercal \Vect{q}
    - \Mat{d}_{\rho \rho} \Mat{\Lambda}_{\rho \rho}^{-1} \Vect{Q}_\rho
    = \Vect{0} .
    \label{eq:spa}
\end{equation}

\noindent When $\Vect{q}$ is evaluated at the ray location $\Vect{q}(\Vect{t})$ in \Eq{eq:MGOpsi}, then, using \Eqs{eq:METrays}, we can simplify
\begin{align}
    \Mat{M}_3 \Mat{R}^\intercal \Vect{q}(\Vect{t})
    &=
    \Mat{M}_3 \Mat{R}^\intercal \Mat{D}^\intercal \Vect{Q}_\Vect{t}(\Vect{t})
    - \Mat{M}_3 \Mat{R}^\intercal \Mat{B}^\intercal \Vect{K}_\Vect{t}(\Vect{t})
    \nonumber\\
    &=
    \Mat{\Lambda}_{\rho \rho}^{-1} \Mat{d}_{\rho \rho}^\intercal \, 
    \Vect{Q}^{\rho}_{\Vect{t}}(\Vect{t})
    - \Vect{K}_\Vect{t}^\rho(\Vect{t})
    ,
\end{align}

\noindent where we have defined the vector projections 
\begin{subequations}
    \begin{align}
        \Mat{L}^\intercal \Vect{K}_\Vect{t}(\Vect{t})
        &\doteq
        \begin{pmatrix}
            \Vect{K}_\Vect{t}^\rho(\Vect{t}) \\[1mm]
            \Vect{K}_\Vect{t}^\varsigma(\Vect{t})
        \end{pmatrix}
        , \\
        \label{eq:QtProj}
        \Mat{L}^\intercal \Vect{Q}_\Vect{t}(\Vect{t})
        &\doteq
        \begin{pmatrix}
            \Vect{Q}_\Vect{t}^\rho(\Vect{t}) \\[1mm]
            \Vect{Q}_\Vect{t}^\varsigma(\Vect{t})
        \end{pmatrix}
        =
        \begin{pmatrix}
            \Vect{Q}_\Vect{t}^\rho(\Vect{t}) \\[1mm]
            \Mat{a}_{\varsigma \varsigma} \Vect{q}_\varsigma(\Vect{t})
        \end{pmatrix}
        .
    \end{align}
\end{subequations}

\noindent Note that \Eq{eq:QtProj} follows from \Eq{eq:METrays}. Since
\begin{align}
    \pd{\Vect{Q}_\rho} \Theta_\Vect{t}
    \left[
        \Mat{L} 
        \begin{pmatrix}
            \Vect{Q}_\rho \\[1mm]
            \Mat{a}_{\varsigma \varsigma} \Vect{q}_\varsigma
        \end{pmatrix} 
    \right]
    &=
    \begin{pmatrix}
        \leftarrow \unit{\Vect{\ell}}_{1}^\intercal \rightarrow \\
        \vdots \\
        \leftarrow \unit{\Vect{\ell}}_{\rho}^\intercal \rightarrow
    \end{pmatrix}
    \Vect{K}_\Vect{t}
    \left[
        \Mat{L} 
        \begin{pmatrix}
            \Vect{Q}_\rho \\[1mm]
            \Mat{a}_{\varsigma \varsigma} \Vect{q}_\varsigma
        \end{pmatrix} 
    \right]
    \nonumber\\
    &\doteq
    \Vect{K}^\rho_\Vect{t} 
    \left[
        \Mat{L} 
        \begin{pmatrix}
            \Vect{Q}_\rho \\[1mm]
            \Mat{a}_{\varsigma \varsigma} \Vect{q}_\varsigma
        \end{pmatrix} 
    \right]
    ,
\end{align}

\noindent the saddlepoint criterion \eq{eq:spa} therefore becomes
\begin{align}
    &\left\{
        \Vect{K}^\rho_\Vect{t} 
        \left[
            \Mat{L} 
            \begin{pmatrix}
                \Vect{Q}_\rho \\[1mm]
                \Mat{a}_{\varsigma \varsigma} \Vect{q}_\varsigma(\Vect{t})
            \end{pmatrix} 
        \right]
        - \Vect{K}_\Vect{t}^\rho(\Vect{t})
    \right\}
    \nonumber\\
    &\hspace{3cm}+ \Mat{d}_{\rho \rho} \Mat{\Lambda}_{\rho \rho}^{-1}
    \left[ 
        \Vect{Q}^{\rho}_{\Vect{t}}(\Vect{t})
        -
        \Vect{Q}_\rho
        \nullFrac
    \right]
    =
    \Vect{0}
    ,
    \label{eq:statPHASE}
\end{align}

\noindent where we have used the fact that $\Mat{d}_{\rho \rho} \Mat{\Lambda}_{\rho \rho}^{-1}$ is symmetric. 

As can be verified, the desired point $\Vect{t}$ on the dispersion manifold is a root to \Eq{eq:statPHASE}, since both terms in brackets vanish simultaneously when $\Vect{Q}_\rho = \Vect{Q}_\Vect{t}^\rho(\Vect{t})$. Let us therefore define the new integration variable
\begin{equation}
    \Vect{\epsilon}_\rho \doteq \Vect{Q}_\rho - \Vect{Q}_\Vect{t}^\rho(\Vect{t})
    ,
    \quad
    \dd \Vect{\epsilon}_\rho = \dd \Vect{Q}_\rho .
\end{equation}

\noindent This yields a modified version of \Eq{eq:MGOpsiT}:
\begin{equation}
    \psi_\Vect{t}(\Vect{q}) 
    = \frac
    {
        \sigma_\Vect{t} \, 
        \alpha_\Vect{t}
        \exp
        \left[
        - \frac{i}{2} \beta_\Vect{t}^\rho(\Vect{q})
        \right]
    }
    {
        (- 2 \pi i)^{\rho/2}
        \sqrt
        {
            \det \Mat{\Lambda}_{\rho \rho}
            \det \Mat{a}_{\varsigma \varsigma}^{-1}
        }
    }
    \, \Upsilon_\Vect{t}^\rho(\Vect{q})
    ,
    \label{eq:MGOpsiTB}
\end{equation}

\noindent where we have defined
\begin{subequations}
    \begin{align}
        \beta_\Vect{t}^\rho (\Vect{q})
        &\doteq
        2 \, \widetilde{g}_\Vect{t}[\Vect{Q}_\Vect{t}^\rho(\Vect{t}), \Vect{q} ]
        ,
        \\
        \Upsilon_\Vect{t}^\rho(\Vect{q})
        &\doteq
        \int_{\cont{0}} \dd \Vect{\epsilon}_\rho \,
        \Psi_\Vect{t}
        \left[
            \Mat{L}
            \begin{pmatrix}
                \Vect{Q}_\Vect{t}^\rho(\Vect{t}) + \Vect{\epsilon}_\rho \\
                \Mat{a}_{\varsigma \varsigma} \Vect{q}_\varsigma
            \end{pmatrix}
        \right]
        \nonumber\\
        &\hspace{29mm}\times
        \exp
        \left[
            - i \gamma_\Vect{t}^\rho(\Vect{\epsilon}_\rho, \Vect{q})
        \right]
        ,
        \\
        \gamma_\Vect{t}^\rho(\Vect{\epsilon}_\rho, \Vect{q})
        &\doteq
        \frac{1}{2} \Vect{\epsilon}_\rho^\intercal \, \Mat{d}_{\rho \rho} \Mat{\Lambda}_{\rho \rho}^{-1} \, \Vect{\epsilon}_\rho
        \nonumber\\
        &\hspace{13mm}+ \Vect{\epsilon}_\rho^\intercal
        \left[
            \Mat{d}_{\rho \rho} \Mat{\Lambda}_{\rho \rho}^{-1} \Vect{Q}_\Vect{t}^\rho(\Vect{t})
            - \Mat{M}_3 \Mat{R}^\intercal \Vect{q}
        \right] 
        .
    \end{align}
\end{subequations}

\noindent Note that $\Upsilon_\Vect{t}^\rho(\Vect{q})$ is integrated along the steepest-descent contour $\cont{0}$ passing through $\Vect{\epsilon}_\rho = \Vect{0}$. Equation \eq{eq:MGOpsi} with $\psi_\Vect{t}$ computed via \Eq{eq:MGOpsiTB} constitutes the generalization of MGO to all values of $\det \Mat{B}$. Thus, MGO can now be applied to any ray pattern in arbitrary media. 


\subsection{Metaplectic geometrical optics with Gaussian coherent states}
\label{sec:coherMGO}

Instead of performing an SVD of $\Mat{B}$, we can develop an expression equivalent to \Eq{eq:MGOpsiTB} using
\begin{align}
    \hspace{-1mm}
    f(\Vect{Q}, \Stroke{\Vect{Z}}_0)
    \doteq
    \exp
    \left[
        - \frac
        {
            |\Vect{Q} - \Vect{Q}_0|^2
        }{2}
        - i \Vect{K}_0^\intercal 
        \left(
            \Vect{Q}
            - \frac{\Vect{Q}_0}{2}
        \right)
    \right]
    .
\end{align}

\noindent These functions, which satisfy the completeness relation
\begin{equation}
    \delta(\Vect{Q} - \Vect{Q}')
    =
    \int \frac{\dd \Vect{Q}_0 \, \dd \Vect{K}_0}{(2\pi)^N \pi^{N/2}}
    \, 
    f(\Vect{Q}, \Stroke{\Vect{Z}}_0)
    \left[
        f(\Vect{Q}', \Stroke{\Vect{Z}}_0)
    \right]^*
    ,
    \label{eq:coherCOMPLETE}
\end{equation}

\noindent can be understood as the spatial representations of the Gaussian coherent states centered around $\Stroke{\Vect{Z}}_0 \doteq (\Vect{Q}_0, \Vect{K}_0)$ in phase space. These states are commonly used in quantum optics~\cite{Scully12} and are discussed in detail in \App{app:Gauss} and supplementary material. As shown in \App{app:Gauss}, the property \eq{eq:coherCOMPLETE} ultimately leads to an alternate representation of the MT:
\begin{align}
    M_\Vect{t}^{-1}(\Vect{q}, \Vect{Q})
    = 
    \int \dd \Vect{K}_0 \,
    \frac
    {
        \sigma \,
        \exp
        \left[
            \fourier{G}_\Vect{t}(\Vect{q}, \Vect{\xi})
            - |\Vect{K}_0|^2
        \right]
    }
    {
        (\sqrt{2} \pi)^N 
        \sqrt{ \det(2 \Mat{D}_\Vect{t} - i \Mat{B}_\Vect{t} ) }
    }
    ,
    \label{eq:gaussMT}
\end{align}

\noindent where we have defined $\Vect{\xi} \doteq \Vect{Q} + 2 i \Vect{K}_0$ and 
\begin{align}
    \hspace{-1mm}\fourier{G}_\Vect{t}(\Vect{q}, \Vect{\xi})
    & \doteq
    - \frac{1}{2} \Vect{q}^\intercal 
    \left(
        2\Mat{D}_\Vect{t} 
        - i \Mat{B}_\Vect{t}
    \right)^{-1}
    \left(
        \Mat{A}_\Vect{t}
        + 2 i \Mat{C}_\Vect{t}
    \right)
    \Vect{q}
    \nonumber\\
    &\hspace{4mm}
    +
    \left(
        \Vect{q}
        - \frac{1}{2} \Mat{D}_\Vect{t}^\intercal \Vect{\xi}
    \right)^\intercal
    \left(
        2 \Mat{D}_\Vect{t}
        - i \Mat{B}_\Vect{t}
    \right)^{-1} \Vect{\xi}
    .
\end{align}

\noindent Note that the complex matrix $2 \Mat{D} - i \Mat{B}$ is always invertible~\cite{Littlejohn87}. 

With \Eq{eq:gaussMT} as the MT kernel, the phase of \Eq{eq:metINV} is stationary where $\Vect{Q}$ and $\Vect{K}_0$ simultaneously satisfy
\begin{subequations}
    \label{eq:gaussSPA}
    \begin{align}
        2i \Mat{D}_\Vect{t}^\intercal
        \left[
            \Vect{K}_\Vect{t}(\Vect{Q}) 
            - \Vect{K}_0
        \right]
        + \Vect{q}
        - \Mat{D}_\Vect{t}^\intercal \Vect{Q}
        + \Mat{B}_\Vect{t}^\intercal \Vect{K}_\Vect{t}(\Vect{Q})
        &= 0
        , \\
        \Vect{q}
        - \Mat{D}_\Vect{t}^\intercal \Vect{Q}
        + \Mat{B}_\Vect{t}^\intercal \Vect{K}_0 
        &= 0 .
    \end{align}
\end{subequations}

\noindent When $\Vect{q}$ is evaluated at $\Vect{q}(\Vect{t})$, a simultaneous solution to \Eqs{eq:gaussSPA} is $\Vect{Q} = \Vect{Q}_\Vect{t}(\Vect{t})$ and $\Vect{K}_0 = \Vect{K}_\Vect{t}(\Vect{t})$. Therefore, upon defining the new integration variables
\begin{align}
    \Vect{\epsilon}_r \doteq \Vect{Q} - \Vect{Q}_\Vect{t}(\Vect{t})
    , \quad
    \Vect{\epsilon}_i \doteq 2 \Vect{K}_0 - 2 \Vect{K}_\Vect{t}(\Vect{t})
    ,
\end{align}

\noindent we obtain the following alternate representation of $\psi_\Vect{t}(\Vect{q})$:
\begin{align}
    &\psi_\Vect{t}(\Vect{q})
    =
    \frac
    {
        \sigma_\Vect{t}
        \alpha_\Vect{t}
        \exp
        \left\{
            \fourier{G}_\Vect{t}[ \Vect{q}, \Vect{\xi}_\Vect{t}(\Vect{t})]
            - |\Vect{K}_\Vect{t}(\Vect{t})|^2 
        \right\}
    }
    {
        (2\sqrt{2} \, \pi)^N 
        \sqrt{ \det(2 \Mat{D}_\Vect{t} - i \Mat{B}_\Vect{t} ) }
    }
    \nonumber\\
    &\hspace{4mm}\times
    \int_{\cont{0}} \dd \Vect{\epsilon}_r \, \dd \Vect{\epsilon}_i \,
    \Psi_\Vect{t}[\Vect{\epsilon}_r + \Vect{Q}_\Vect{t}(\Vect{t})]
    \exp
    \left[
        - \fourier{\gamma}_\Vect{t}(\Vect{\epsilon}, \Vect{q}, \Vect{t})
    \right]
    ,
    \label{eq:MGOpsiGAUSS}
\end{align}

\noindent where we have defined
\begin{subequations}
    \begin{align}
        \Vect{\epsilon} &\doteq \Vect{\epsilon}_r + i \Vect{\epsilon}_i
        , \\
        \Vect{\xi}_\Vect{t}(\Vect{t}) &\doteq \Vect{Q}_\Vect{t}(\Vect{t}) + 2 i \Vect{K}_\Vect{t}(\Vect{t})
        , \\
        \fourier{\gamma}_\Vect{t}(\Vect{\epsilon}, \Vect{q}, \Vect{t})
        &\doteq
        \frac{1}{2} \Vect{\epsilon}^\intercal \Mat{D}_\Vect{t}
        \left(
            2 \Mat{D}_\Vect{t}
            - i \Mat{B}_\Vect{t}
        \right)^{-1} \Vect{\epsilon}
        + \frac
        {
            |\Vect{\epsilon}_i|^2
        }{4}
        + \Vect{\epsilon}_i^\intercal \Vect{K}_\Vect{t}(\Vect{t})
        \nonumber\\
        &\hspace{4mm}
        - \Vect{\epsilon}^\intercal 
        \left(
            2 \Mat{D}_\Vect{t}
            - i \Mat{B}_\Vect{t}
        \right)^{-\intercal}
        \left[
            \Vect{q}
            - \Mat{D}_\Vect{t}^\intercal \Vect{\xi}_\Vect{t}(\Vect{t})
        \right]
        .
    \end{align}
\end{subequations}

Equation \eq{eq:MGOpsiGAUSS} is equivalent to \Eq{eq:MGOpsiTB} but might be advantageous since it can be applied `as is' without performing an SVD of $\Mat{B}_\Vect{t}$. That said, \Eq{eq:MGOpsiGAUSS} involves a $2N$-D integral, which is harder to evaluate numerically. In this sense, the representation \eq{eq:MGOpsiTB} may be more practical, especially at large $N$.


\section{Examples}
\label{sec:ex}

Here, we consider several examples of MGO, which have been shortened for clarity. The complete calculations can be found in the supplementary material.


\subsection{Plane wave in uniform medium: No caustic}

As a first example, let us consider a plane wave propagating in a uniform medium. For simplicity, we consider $1$-D propagation governed by the one-way wave equation
\begin{equation}
    i\pd{q} \psi(q) + \psi(q) = 0
    .
    \label{eq:ex1Wave}
\end{equation}

\noindent There is no caustic in this case, and \Eq{eq:ex1Wave} is easy to integrate even without using MGO. However, this example is instructive to illustrate the reformulated MGO machinery when $\det{\Mat{B}_\Vect{t}} = 0$ with relatively little algebra.

Let us start by writing \Eq{eq:ex1Wave} in the integral form \eq{eq:WAVEintegral}. The corresponding kernel $D(q, q')$ can be written as
\begin{equation}
    D(q, q') 
    = 
    i \pd{q'} \delta(q - q') 
    - \delta(q - q') ,
\end{equation}

\noindent so the Weyl symbol \eq{eq:Wigner} is as follows:
\begin{equation}
    \Symb{D}(q, k) = k - 1 .
\end{equation}

\noindent The corresponding ray equations are
\begin{equation}
    \pd{\tau} q(\tau) = 1 
    , \quad
    \pd{\tau} k(\tau) = 0
    ,
    \label{eq:ex1RayEQ}
\end{equation}

\noindent with solutions given by
\begin{equation}
    q(\tau) = \tau 
    , \quad
    k(\tau) = 1
    ,
\end{equation}

\noindent where the integration constants have been chosen to satisfy $\Symb{D}[q(0), k(0)] = 0$. Since $\tau(q) = q$ is single-valued, $\psi(q)$ will be absent of caustics.

From \Eq{eq:ex1RayEQ}, it is clear that the tangent plane of the dispersion manifold is $\Vect{q}$-space itself, that is,
\begin{equation}
    \Mat{S}_t = \IMat{2} 
\end{equation}

\noindent for all $t \in \tau(q)$. Hence, $\Psi_t(Q_t)$ is trivially obtained as
\begin{equation}
    \Psi_t(Q_t) = \alpha_t \exp(i Q_t - i t)
    ,
\end{equation}

\noindent where we have used $Q_t(t) = t$. Since $\pd{t}\Mat{S}_t = \OMat{2}$, $\pd{Q}\Psi_t(Q) = 0$, and $\pd{t}Q_t(t) = 1$, we also compute
\begin{equation}
    \eta_t = i 
    , \quad
    \alpha_t = \alpha_0 \exp\left( i t \right)
    ,
\end{equation}

\noindent where $\alpha_0$ is a constant. Then, since $\Mat{B}_t = 0$ implies that $\rho = 0$, $\varsigma = 1$, and $\Mat{R} = \Mat{L} = 1$, then $\beta_t^\rho = \gamma_t^\rho = 0$ and the integration over $\dd \epsilon_\rho$ is empty. Hence, \Eq{eq:MGOpsiTB} becomes
\begin{equation}
    \psi_t[q(t)] = \sigma_t \alpha_0 \exp[i Q_t(t)] .
    \label{eq:ex1PsiT}
\end{equation}

\noindent Since the branch cut of the MT is never crossed, we can take $\sigma_t = 1$. Then, the summation over branches is trivially performed to yield
\begin{equation}
    \psi(q) = \sum_{t \in \tau(q)} \psi_t[q(t)] = \alpha_0 \exp\left(i q \right) .
    \label{eq:planeSOL}
\end{equation}

\noindent Equation \eq{eq:planeSOL} is an exact solution of \Eq{eq:ex1Wave}, which is anticipated because \eq{eq:ex1Wave} is a first-order equation and thus coincides with its GO approximation.


\subsection{Plane wave in linearly stratified medium: Fold caustic}

As a second example, let us consider oblique propagation in a linearly stratified medium. Suppose that the wave is described by the Helmholtz-type equation
\begin{equation}
    \pd{\Vect{q}}^2 \psi( \Vect{q} ) 
    + (k_0^2 - q_1) \psi( \Vect{q} ) = 0
    ,
    \label{eq:airyEQ}
\end{equation}

\noindent where $k_0$ is a constant and $\pd{\Vect{q}}^2 \doteq \pd{q_1}^2 + \pd{q_2}^2$. Our coordinate system is such that $q_1$ is aligned with the medium stratification and $q_2$ is transverse to $q_1$. Let us also consider the initial condition
\begin{equation}
    \psi(0, q_2) = c \exp(i k_0 q_2) ,
    \label{eq:airyIC}
\end{equation}

\noindent where $c$ is an arbitrary constant.

Equation \eq{eq:airyEQ} can be equivalently written as an integral equation \eq{eq:WAVEintegral} with integration kernel
\begin{equation}
    D(\Vect{q}, \Vect{q}') 
    = 
    - \pd{\Vect{q}'}^2 \delta(\Vect{q} - \Vect{q}') 
    + (q_1 - k_0^2) \delta(\Vect{q} - \Vect{q}') .
\end{equation}

\noindent The corresponding Weyl symbol \eq{eq:Wigner} is
\begin{equation}
    \Symb{D}(\Vect{q}, \Vect{k}) 
    =
    k_1^2 + k_2^2 + q_1 - k_0^2
    ,
\end{equation}

\noindent and the corresponding ray equations are
\begin{subequations}
    \begin{align}
        \pd{\tau_1} q_1(\Vect{\tau}) &= 2 k_1(\Vect{\tau})
        , \quad
        \pd{\tau_1} k_1(\Vect{\tau}) = -1
        , \\
        \pd{\tau_1} q_2(\Vect{\tau}) &= 2 k_2(\Vect{\tau})
        , \quad
        \pd{\tau_1} k_2(\Vect{\tau}) = 0
        .
    \end{align}
\end{subequations}

\noindent Let us define $\tau_1$ such that $q_1(0, \tau_2) = 0$. Then, the initial condition \eq{eq:airyIC} implies that $k_2(0, \tau_2) = k_0$, and the local dispersion relation $\Symb{D}[\Vect{q}(0, \tau_2), \Vect{k}(0, \tau_2)] = 0$ requires $k_1(0, \tau_2) = 0$. This leaves $q_2(0, \tau_2)$ undetermined. Since $\tau_2$ must parameterize the initial conditions for the rays, let us choose $q_2(0, \tau_2) = \tau_2$. Hence, the ray solutions are
\begin{subequations}
    \begin{align}
        q_1(\Vect{\tau}) 
        &= - \tau_1^2
        , \quad
        q_2(\Vect{\tau}) 
        = \tau_2 + 2 k_0 \tau_1
        , \\
        k_1( \Vect{\tau}) 
        &=  - \tau_1 
        , \quad
        k_2( \Vect{\tau}) 
        =  k_0
        .
    \end{align}
\end{subequations}

\noindent The inverse function $\Vect{\tau}(\Vect{q})$ is calculated as
\begin{equation}
    \tau_1(\Vect{q}) = \pm \sqrt{- q_1} 
    , \quad
    \tau_2(\Vect{q}) = q_2 \mp 2 k_0 \sqrt{- q_1}
    .
    \label{eq:airyBRANCH}
\end{equation}

\noindent Clearly, $\Vect{\tau}(\Vect{q})$ is double-valued, so there are two branches that must ultimately be summed over.

A basis for $\Vect{Q}_\Vect{t}$-space is provided by the vector pair
\begin{subequations}
    \begin{align}
        \pd{\tau_1} \Vect{z}( \Vect{t})
        &=
        \begin{pmatrix}
            - 2 t_1
            &
            2 k_0
            &
            -1
            &
            0
        \end{pmatrix}^\intercal
        , \\
        \pd{\tau_2} \Vect{z}( \Vect{t})
        &=
        \begin{pmatrix}
            0
            &
            1
            &
            0
            &
            0
        \end{pmatrix}^\intercal
        ,
    \end{align}
\end{subequations}

\noindent where $\Vect{z}(\Vect{t}) \doteq (\Vect{q}(\Vect{t}), \Vect{k}(\Vect{t}))^\intercal$. Then, symplectic Gram--Schmidt orthogonalization~\cite{Lopez20a} yields the submatrices
\begin{subequations}
    \begin{align}
        \Mat{A}_\Vect{t} 
        &= \Mat{D}_\Vect{t} =
        \frac{1}{\vartheta_\Vect{t}}
        \begin{pmatrix}
            - 2 t_1 & 0 \\
            0 & \vartheta_\Vect{t}
        \end{pmatrix}
        , \\
        \Mat{B}_\Vect{t} 
        &= - \Mat{C}_\Vect{t} = 
        \frac{1}{\vartheta_\Vect{t}}
        \begin{pmatrix}
            - 1 & 0 \\
            0 & 0
        \end{pmatrix}
        ,
    \end{align}
\end{subequations}

\noindent where we have defined
\begin{equation}
    \vartheta_\Vect{t} \doteq \sqrt{1 + 4 t_1^2}
    .
\end{equation}

The rotated rays are calculated via \Eq{eq:METrays} as
\begin{align}
    \Vect{Q}_\Vect{t}(\Vect{\tau})
    =
    \begin{pmatrix}
        \frac
        {
            2 t_1 \tau_1^2 + \tau_1
        }{\vartheta_\Vect{t} }
        \\
        \tau_2 + 2 k_0 \tau_1 
    \end{pmatrix}
    , \quad
    \Vect{K}_\Vect{t}(\Vect{\tau}) 
    = 
    \begin{pmatrix}
        \frac
        {
            2 t_1 \tau_1 - \tau_1^2
        }{\vartheta_\Vect{t} }
        \\
        k_0
    \end{pmatrix}
    .
\end{align}

\noindent We can therefore compute the inverse function $\Vect{\tau}(\Vect{Q}_\Vect{t})$, which is double-valued. Upon restricting $\Vect{K}_\Vect{t}(\Vect{Q}_\Vect{t})$ to the correct branch, we obtain
\begin{subequations}
    \begin{align}
        K_{\Vect{t}, 1}(\Vect{Q}_\Vect{t}) 
        &= 
        - \frac{Q_{\Vect{t},1}}{2 t_1}
        - \vartheta_\Vect{t}
        \frac
        {
            1 
            - \sqrt{1 + 8  t_1 \vartheta_\Vect{t} Q_{\Vect{t},1}}
        }{8 t_1^2}
        , \\
        K_{\Vect{t}, 2}(\Vect{Q}_\Vect{t})
        &= k_0
        .
    \end{align}
\end{subequations}

\noindent Then, the line integral of \Eq{eq:MGOphase} is computed to yield
\begin{align}
    \Theta_\Vect{t}[\Vect{\epsilon} + \Vect{Q}_\Vect{t}(\Vect{t})]
    &=
    k_0 \epsilon_2
    + \frac{8 t_1^4 - \vartheta^4_\Vect{t} }{8 t_1^2 \vartheta_\Vect{t}} \epsilon_1
    - \frac{1}{4 t_1} \epsilon_1^2
    \nonumber\\
    &\hspace{4mm}+ \frac
    {
        \left(\vartheta^4_\Vect{t} + 8 t_1 \vartheta_\Vect{t} \epsilon_1\right)^{3/2}
        - \vartheta^6_\Vect{t}
    }{96 t_1^3}
    .
\end{align}

\noindent Using \Eqs{eq:MGOphi} and \eq{eq:MGOjac}, we next compute
\begin{equation}
    \Phi_\Vect{t}
    \left[
        \Vect{\epsilon}
        + \Vect{Q}_\Vect{t}(\Vect{t})
    \right]
    =
    \frac
    {
        \vartheta_\Vect{t}
    }
    {
        \left( 
            \vartheta^4_\Vect{t}
            + 8 t_1 \vartheta_\Vect{t} \epsilon_1 
        \right)^{1/4}
    }
    ,
\end{equation}

\noindent where we have chosen the initial conditions to satisfy
\begin{equation}
    \Phi_\Vect{t}\left[\Vect{Q}_\Vect{t}(t_1, \tau_2)\right]
    = 1
    .
\end{equation}

\noindent We then compute via \Eqs{eq:MGOalpha} and \eq{eq:MGOeta}
\begin{align}
    \alpha_\Vect{t}
    =
    \frac{\alpha_{\left(0, t_2 \right)}}{\sqrt{ \vartheta_\Vect{t}} }
    \exp
    \left(
        2 i k_0^2 t_1
        + i \frac{2 t_1^3}{3}
        - i \frac{t_1^5}{\vartheta^2_\Vect{t}}
    \right)
    ,
\end{align}

\noindent where $\alpha_{\left(0, t_2 \right)}$ is an arbitrary initial condition.

We now perform the inverse MT. Note that $\Mat{B}_\Vect{t}$ is already in the desired SVD form, with $\Mat{L} = \Mat{R} = \IMat{2}$ and $\rho = 1$. Since $\Vect{\epsilon}_\rho = \epsilon_1$ and $\Vect{Q}_\Vect{t}^\rho(\Vect{t}) = Q_{\Vect{t}, 1}(\Vect{t})$, we compute
\begin{equation}
    \beta_\Vect{t}^\rho [\Vect{q}(\Vect{t})]
    =
    - \frac{2 t_1^5}{\vartheta^2_\Vect{t}}
    , \quad
    \gamma_\Vect{t}^\rho [\Vect{\epsilon}_\rho, \Vect{q}(\Vect{t})]
    =
    \frac{t_1^2}{\vartheta_\Vect{t}} \epsilon_1
    + t_1 \epsilon_1^2 
    .
\end{equation}

\noindent Hence, we obtain
\begin{align}
    &\Upsilon_\Vect{t}^\rho \left[\Vect{q}(\Vect{t}) \right]
    =
    \int_{\cont{0}} \dd \epsilon_1 \,
    \frac
    {
        \vartheta_\Vect{t}
    }
    {
        \left( 
            \vartheta^4_\Vect{t} 
            + 8 t_1 \vartheta_\Vect{t} \epsilon_1 
        \right)^{1/4}
    }
    \nonumber\\
    &\times\exp
    \left[
        i \frac
        {
            \left(\vartheta^4_\Vect{t} + 8 t_1 \vartheta_\Vect{t} \epsilon_1\right)^{3/2}
            - \vartheta^6_\Vect{t}
        }{96 t_1^3}
        - i \frac{\vartheta^2_\Vect{t}}{4 t_1} \epsilon_1^2
        - i \frac{\vartheta^3_\Vect{t} }{8 t_1^2} \epsilon_1
    \right]
    .
\end{align}

\noindent This is the same integral that was studied in \Ref{Lopez20a}, where the following approximation was derived:
\begin{align}
    \Upsilon_\Vect{t}^\rho \left[\Vect{q}(\Vect{t}) \right]
    &\approx
    \pi \vartheta_\Vect{t} \exp\left(- i \frac{2}{3} t_1^3 \vartheta^6_\Vect{t} \right)
    \nonumber\\
    &\hspace{4mm}\times
    \left[
        \airyA
        \left(
            - t_1^2 \vartheta^4_\Vect{t} 
        \right) 
        - i \, \frac{t_1}{|t_1|} 
        \airyB
        \left(
            - t_1^2 \vartheta^4_\Vect{t} 
        \right) 
    \right]
    ,
\end{align}

\noindent where $\airyA(x)$ and $\airyB(x)$ are the Airy functions of the first and second kind, respectively~\cite{Olver10a}. Thus, \Eq{eq:MGOpsiTB} yields
\begin{align}
    \psi_\Vect{t}[\Vect{q}(\Vect{t})]
    &=
    i 
    \sigma_\Vect{t}
    \alpha_{(0, t_2)}
    \frac
    {
        \vartheta_\Vect{t}
        \sqrt{\pi}
    }
    {
        \sqrt{- 2 i}
    }
    \exp
    \left[
        2 i k_0^2 t_1
        + i \frac{2 t_1^3}{3} \left(1 - \vartheta^6_\Vect{t} \right)
    \right]
    \nonumber\\
    &\hspace{4mm}\times
    \left[
        \airyA
        \left(
            - t_1^2 \vartheta^4_\Vect{t} 
        \right) 
        - i \, \frac{t_1}{|t_1|} 
        \airyB
        \left(
            - t_1^2 \vartheta^4_\Vect{t} 
        \right) 
    \right]
    .
\end{align}

\begin{figure}
    \centering
    \includegraphics[width=\linewidth,trim={2mm 16mm 2mm 14mm},clip]{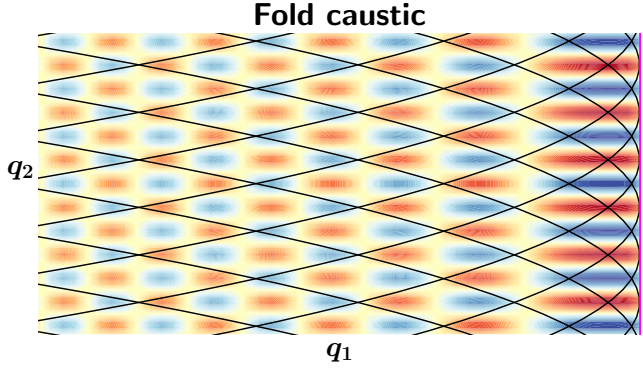}
    \caption{Contour plot showing the real part of the MGO solution \eq{eq:airyMGO}, with $k_0 = 2$, near the fold caustic (cutoff) located at $q_1 = 0$ (magenta). The ray trajectories $( q_1(\Vect{\tau}), q_2(\Vect{\tau}) )$ are shown as black curves. Note that the field remains finite along the caustic. In fact, the MGO solution is nearly indistinguishable with the exact solution \eq{eq:airyEXACT}.}
    \label{fig:airy}
\end{figure}

Since the MT branch cut is never crossed, we can take $\sigma_\Vect{t} = 1$. Then, summing over both branches of $\Vect{\tau}(\Vect{q})$ [\Eq{eq:airyBRANCH}] and choosing
\begin{equation}
    \alpha_{(0, t_2)} 
    = \frac{\sqrt{-2 i}}{2i \sqrt{\pi}} \exp( i k_0 t_2)
\end{equation}

\noindent to satisfy the initial condition \eq{eq:airyIC} ultimately yields
\begin{align}
    \psi(\Vect{q}) 
    &=
    \sqrt{1 - 4 q_1} \,
    \exp(i k_0 q_2)
    \nonumber\\
    &\hspace{4mm}\times
    \left\{
        \airyA[- \varrho^2(q_1) ]
        \cos \varpi(q_1) 
        \right.\nonumber\\
        &\left.\hspace{27mm}
        -
        \airyB[- \varrho^2(q_1) ]
        \sin \varpi(q_1) 
    \right\}
    ,
    \label{eq:airyMGO}
\end{align}

\noindent where we have chosen $c = \airyA( 0 )$ and defined
\begin{subequations}
    \begin{align}
        \varrho(q_1) &\doteq (1 - 4 q_1) \sqrt{-q_1}
        , \\
        \varpi(q_1) &\doteq \frac{2}{3} \varrho^3(q_1) - \frac{2}{3} (-q)^{3/2}
    .
    \end{align}
\end{subequations}

\noindent The MGO solution \eq{eq:airyMGO} is plotted in \Fig{fig:airy}. Notably, this solution is finite along the caustic surface (cutoff) located at $q_1 = 0$, and agrees remarkably well with the exact solution of \Eq{eq:airyEQ},
\begin{equation}
    \psi_\text{ex}(\Vect{q}) = \airyA(q_1) \exp(ik_0 q_2)
    .
    \label{eq:airyEXACT}
\end{equation}

\noindent A similar plot of \Eq{eq:airyEXACT} is not presented because it is virtually indistinguishable from \Fig{fig:airy}.


\subsection{Imperfectly focused plane wave in uniform medium: Cusp caustic}

\begin{figure*}
    \centering
    \includegraphics[width=0.3\linewidth]{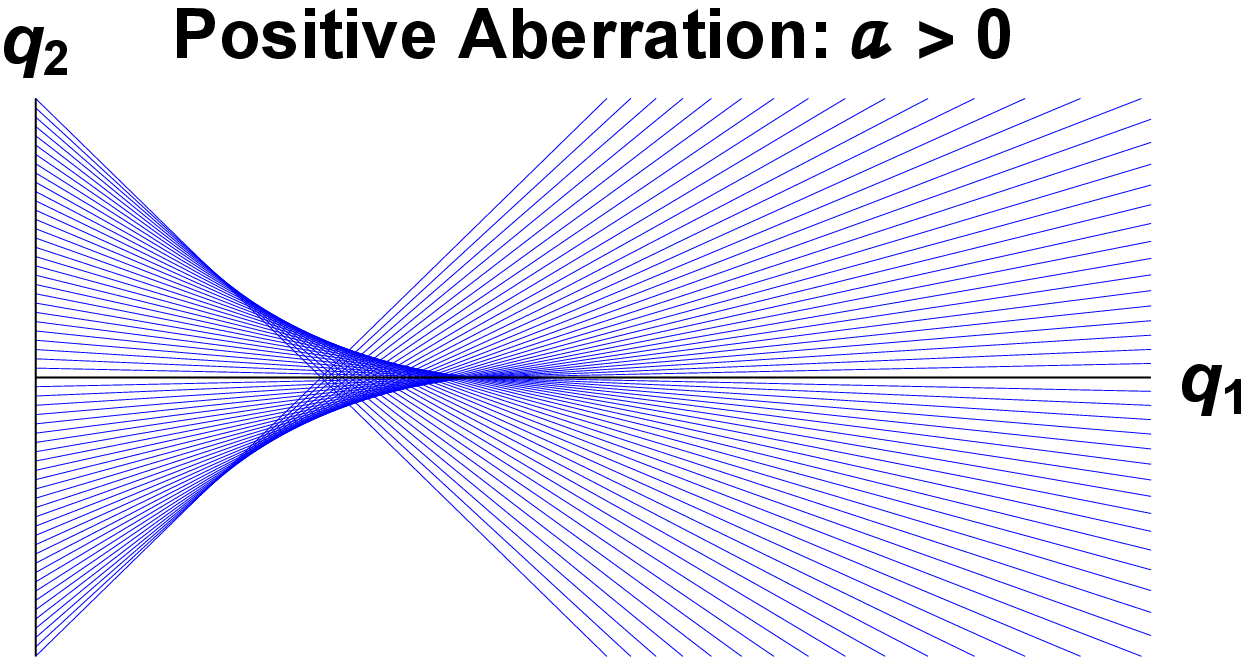}
    \hspace{3mm}
    \includegraphics[width=0.3\linewidth]{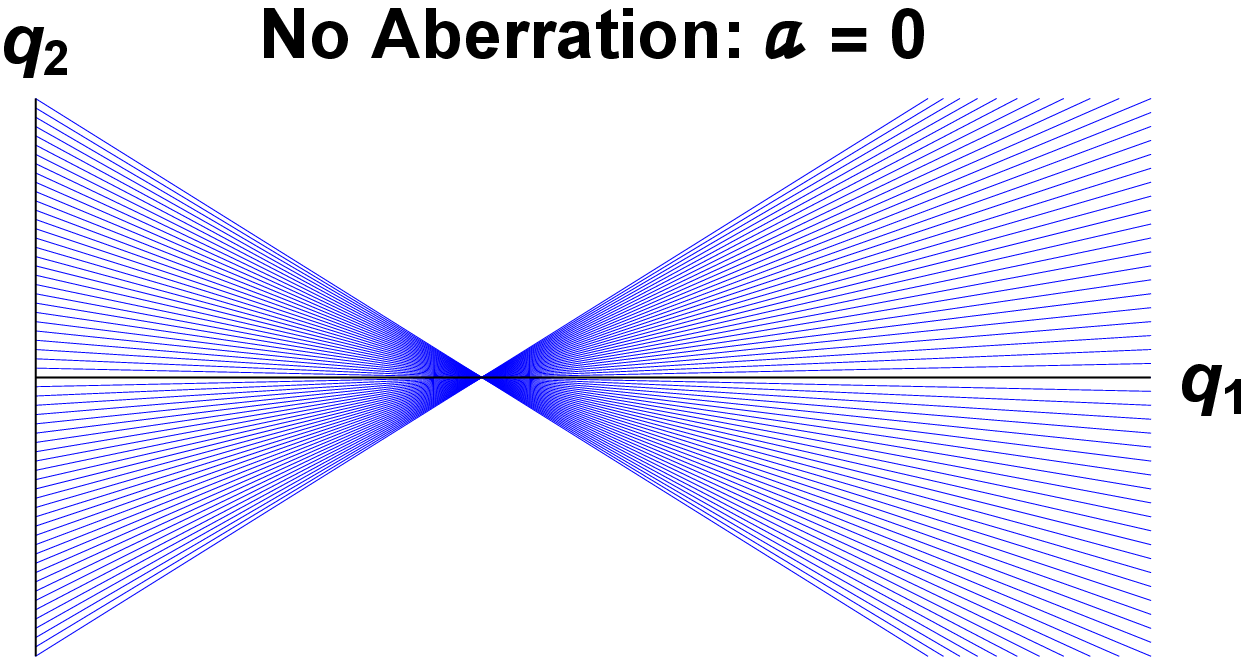}
    \hspace{3mm}
    \includegraphics[width=0.3\linewidth]{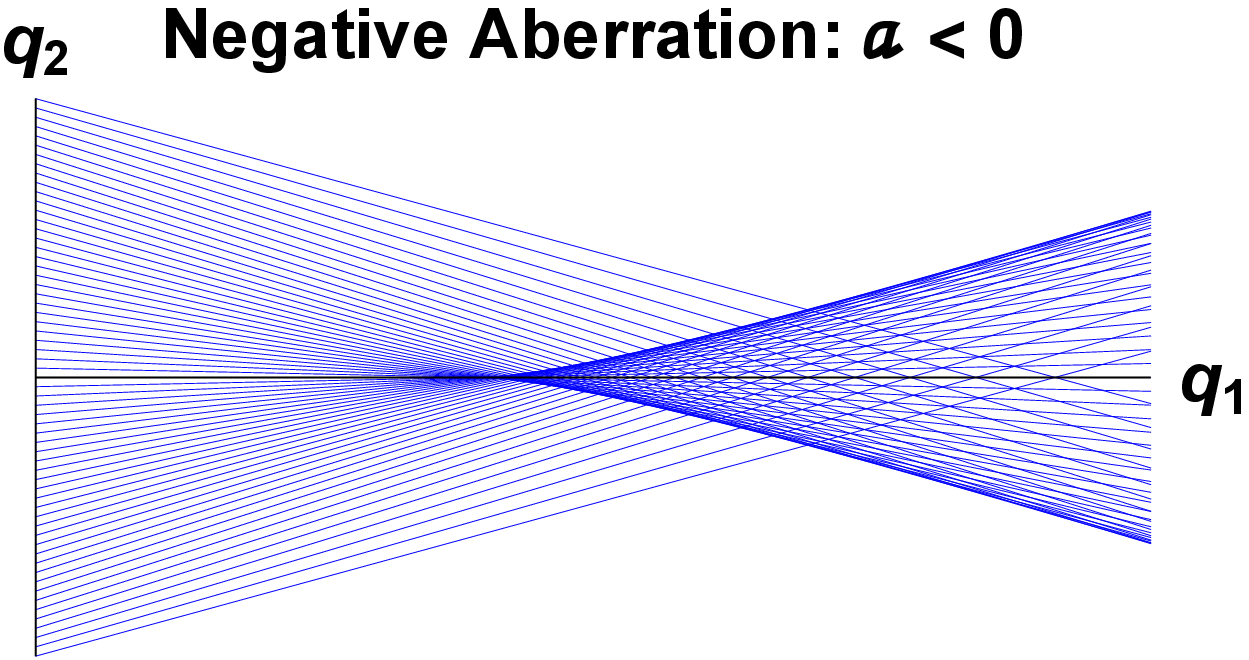}
    \caption{Ray trajectories for the paraxial equation \eq{eq:cuspEQ} with initial conditions given by \Eq{eq:cuspIC}. For no aberration ($\aber = 0$), the rays focus at $\Vect{q} = (f, 0)$. Positive aberration ($\aber > 0$) causes the outer rays to focus before the focal point, while negative aberration ($\aber < 0$) causes the outer rays to focus beyond the focal point. Both cases result in a cusped ray pattern.}
    \label{fig:aber}
\end{figure*}

As a final example, let us consider a 2-D plane wave described by the paraxial wave equation~\cite{Kogelnik66}
\begin{equation}
    i \pd{q_1} \psi(\Vect{q}) + \frac{1}{2}\pd{q_2}^2 \psi(\Vect{q}) + \psi(\Vect{q}) = 0 ,
    \label{eq:cuspEQ}
\end{equation}

\noindent where $q_1$ is aligned with the optical axis and $q_2$ is tranverse to it. Let us assume the initial condition 
\begin{equation}
    \psi(0 , q_2) = \sqrt{ \frac{2 \pi i }{f } } \exp\left( - \frac{i}{2f} \, q_2^2 - \frac{i \aber}{4 f^3} \, q_2^4 \right) .
    \label{eq:cuspIC}
\end{equation}

\noindent This corresponds to a wave that is being focused by an imperfect lens, with focal distance $f$ and aberration $a$. In the absence of aberration ($\aber = 0$), the initial field will focus at $\Vect{q} = (f, 0)$; however, as shown in \Fig{fig:aber}, finite aberration causes the focusing to become distorted, resulting in a cusped wavefield. For simplicity, we shall assume that $f \gg 1$ and $\aber < 0$.

Equation \eq{eq:cuspEQ} can be equivalently written as an integral equation \eq{eq:WAVEintegral} with integration kernel
\begin{equation}
    D(\Vect{q}, \Vect{q}') 
    = 
    i \pd{q_1'} \delta(\Vect{q} - \Vect{q}') 
    - \frac{1}{2} \pd{q_2'}^2 \delta(\Vect{q} - \Vect{q}')
    - \delta(\Vect{q} - \Vect{q}') .
\end{equation}

\noindent The corresponding Weyl symbol \eq{eq:Wigner} is
\begin{equation}
    \Symb{D}(\Vect{q}, \Vect{k}) 
    =
    k_1 + \frac{k_2^2}{2} - 1
    ,
\end{equation}

\noindent and the corresponding ray equations are
\begin{subequations}
    \label{eq:cuspRAYeq}
    \begin{align}
        \pd{\tau_1} k_1(\Vect{\tau}) = 0
        , \quad
        \pd{\tau_1} q_1(\Vect{\tau}) &= 1
        , \\
        \pd{\tau_1} k_2(\Vect{\tau}) = 0
        , \quad
        \pd{\tau_1} q_2(\Vect{\tau}) &= k_2(\Vect{\tau})
        .
    \end{align}
\end{subequations}

Similar to the previous example, let us define $\tau_1$ and $\tau_2$ such that $q_1(0, \tau_2) = 0$ and $q_2(0, \tau_2) = f \tau_2$. Then, imposing the local dispersion relation \eq{eq:GO1} and the initial condition \eq{eq:cuspIC} yields the ray trajectories
\begin{subequations}
    \begin{align}
        q_1(\Vect{\tau}) &= \tau_1 , \quad
        q_2(\Vect{\tau}) = f \tau_2 + k_2(\Vect{\tau}) \tau_1 ,\\
        k_1(\Vect{\tau}) &=  1 - \frac{ k_2^2(\Vect{\tau})}{2} , \quad
        k_2(\Vect{\tau}) = - \, \tau_2 - \aber \, \tau_2^3 .
    \end{align}
\end{subequations}

\noindent The inverse function $\Vect{\tau}(\Vect{q})$ is either single- or triple-valued, depending on the value of the discriminant
\begin{equation}
    \Delta(\Vect{q}) = 
    4 \left(\frac{q_1 - f}{ \aber q_1} \right)^3 
    + 27 \left( \frac{q_2}{ \aber q_1} \right)^2
    .
\end{equation}

\noindent If $\Delta(\Vect{q}) > 0$, there is only one ray given by
\begin{equation}
    \tau_2^{(0)}(\Vect{q}) = 
    \sqrt[\textrm{\small 3}]{
        - \frac{q_2}{2 \aber q_1} 
        + \sqrt
        {
            \frac{\Delta(\Vect{q})}{108}
        } 
    } 
    + 
    \sqrt[\textrm{\small 3}]{ 
        - \frac{q_2}{2 \aber q_1}  
        - \sqrt
        {
            \frac{\Delta(\Vect{q})}{108}
        } 
    }
    ,
\end{equation}

\noindent while if $\Delta(\Vect{q}) \le 0$, there are two additional rays given by
\begin{equation}
    \tau_2^{(\pm)}(\Vect{q}) = 
    \textrm{Re}
    \left[
        (-1 \pm i \sqrt{3})
        \sqrt[\textrm{\small 3}]{
            - \frac{q_2}{2 \aber q_1} 
            + i \sqrt
            {
                \frac{|\Delta(\Vect{q})|}{108}
            } 
        }
    \right]
    .
\end{equation}

\noindent For all values of $\Delta(\Vect{q})$, one has $\tau_1(\Vect{q}) = q_1$.

A basis for $\Vect{Q}_\Vect{t}$-space is provided by the vector pair
\begin{subequations}
    \begin{align}
        \pd{\tau_1} \Vect{z}( \Vect{t})
        &=
        \begin{pmatrix}
            1
            &
            k_2( \Vect{t} )
            &
            0
            &
            0
        \end{pmatrix}^\intercal
        , \\
        \pd{\tau_2} \Vect{z}( \Vect{t})
        &=
        \begin{pmatrix}
            0
            &
            j_\Vect{t}
            &
            - k_2(\Vect{t}) k_2'(\Vect{t})
            &
            k_2'(\Vect{t})
        \end{pmatrix}^\intercal
        ,
    \end{align}
\end{subequations}

\noindent where we have defined
\begin{equation}
    j_\Vect{t}
    \doteq
    \det \pd{\Vect{\tau}} \Vect{q}(\Vect{t}) 
    = f + t_1 k_2'(\Vect{t})
    .
\end{equation}

\noindent Then, $\Mat{S}_\Vect{t}$ is constructed using symplectic Gram--Schmidt orthogonalization~\cite{Lopez20a}. This yields the submatrices
\begin{subequations}
    \begin{align}
        \Mat{A}_\Vect{t} 
        &= \Mat{D}_\Vect{t}
        = \frac{1}{\vartheta_\Vect{t} \varphi_\Vect{t}}
        \begin{pmatrix}
            \varphi_\Vect{t} & \varphi_\Vect{t} k_2(\Vect{t}) \\
            - j_\Vect{t} k_2(\Vect{t}) & j_\Vect{t}
        \end{pmatrix}
        , \\
        \Mat{B}_\Vect{t} 
        &= - \Mat{C}_\Vect{t}
        = \frac{\vartheta_\Vect{t} k_2'(\Vect{t}) }{\varphi_\Vect{t}}
        \begin{pmatrix}
            0 & 0 \\
            - k_2(\Vect{t}) & 1
        \end{pmatrix}
        ,
    \end{align}
\end{subequations}

\noindent where we have defined
\begin{equation}
    \vartheta_\Vect{t} \doteq \sqrt{1 + k_2^2(\Vect{t})}
    , \quad
    \varphi_\Vect{t} \doteq \sqrt{j_\Vect{t}^2 + \left[k_2'(\Vect{t}) \vartheta^2_\Vect{t} \right]^2} 
    .
\end{equation}

The rotated rays are computed using \Eqs{eq:METrays}, although the result is quite lengthy and will not be shown. Next, using \Eq{eq:MGOphi} we compute the envelope as
\begin{equation}
    \Phi_\Vect{t}(\Vect{\tau})
    =
    \frac
    {
        \varphi_\Vect{t}
    }
    {
        \sqrt
        {
            j_\Vect{t} j_\Vect{\tau} 
            + [1 + k_2(\Vect{t}) k_2(\Vect{\tau})]^2 
            k_2'(\Vect{\tau}) k_2'(\Vect{t})
        }
    }
    ,
\end{equation}

\noindent where $\Phi_\Vect{t}(\Vect{\tau}) \doteq \Phi_\Vect{t}[ \Vect{Q}_\Vect{t}(\Vect{\tau}) ]$ and we have chosen
\begin{equation}
    \Phi_\Vect{t}\left[\Vect{Q}_\Vect{t}(t_1, \tau_2)\right]
    = 1
    .
\end{equation}

Although $\Vect{\tau}(\Vect{Q}_\Vect{t})$ is impractical to construct explicitly, $\pd{\Vect{Q}} \Phi_\Vect{t}$ can still be calculated from $\Phi_\Vect{t}(\Vect{\tau})$ using
\begin{equation}
    \pd{\Vect{Q}} \Phi_\Vect{t}[\Vect{Q}_\Vect{t}(\Vect{t})]
    =
    \left[
        \pd{\Vect{\tau}} \Vect{Q}_\Vect{t}(\Vect{t}) 
    \right]^{-1}
    \pd{\Vect{\tau}} \Phi_\Vect{t}(\Vect{t})
    .
\end{equation}

\noindent Ultimately, \Eqs{eq:MGOalpha} and \eq{eq:MGOeta} yield
\begin{align}
    \alpha_\Vect{t} 
    &= \alpha_{(0,t_2)} \sqrt{ \frac{\varphi_{(0,t_2)} }{\varphi_\Vect{t}} }
    \, \exp
    \left\{
        i t_1 \frac{1 + \vartheta_\Vect{t}^2}{2}
        + \frac{i}{2} \beta_\Vect{t}^\rho [\Vect{q}(\Vect{t})]
        \right. \nonumber\\
        &\left.\hspace{35mm}
        - \frac{i}{2} \beta_{(0, t_2)}^\rho [\Vect{q}(0, t_2)]
    \right\}
    ,
\end{align}

\noindent where $\alpha_{(0,t_2)}$ is determined by the initial conditions and $\beta_\Vect{t}^\rho [\Vect{q}(\Vect{t})]$ is defined below [\Eq{eq:cuspBETA}].

We next perform an SVD of $\Mat{B}_\Vect{t}$ to obtain
\begin{equation}
    \Mat{L}
    =
    \begin{pmatrix}
        0 & 1 \\
        -1 & 0
    \end{pmatrix}
    , \quad
    \Mat{R}
    =
    \frac{1}{\vartheta_\Vect{t}}
    \begin{pmatrix}
        k_2(\Vect{t}) & 1 \\
        - 1 & k_2(\Vect{t})
    \end{pmatrix}
    .
\end{equation}

\noindent Hence, we can compute%
\begin{subequations}
    \begin{gather}
    \label{eq:cuspBETA}
    \beta_\Vect{t}^\rho [\Vect{q}(\Vect{t})]
    =
    \frac{f^2 j_\Vect{t} t_2^2}{\vartheta_\Vect{t}^4 k_2'(\Vect{t})}
    - \frac{2 f \varphi_\Vect{t} t_2 \, Q_{\Vect{t}, 2}(\Vect{t})}{\vartheta_\Vect{t}^3 k_2'(\Vect{t})}
    + \frac{j_\Vect{t} \, Q_{\Vect{t}, 2}^2(\Vect{t})
    }{\vartheta_\Vect{t}^2 k_2'(\Vect{t})}
    , \\
    \gamma_\Vect{t}^\rho[\epsilon_\rho, \Vect{q}(\Vect{t}) ]
    =
    \frac{j_\Vect{t}}{2 \vartheta_\Vect{t}^2 k_2'(\Vect{t})} \epsilon_\rho^2
    - K_{\Vect{t}, 2}(\Vect{t}) \epsilon_\rho
    .
\end{gather}
\end{subequations}

\noindent Thus, \Eq{eq:MGOpsiTB} yields
\begin{align}
    \psi_\Vect{t}[\Vect{q}(\Vect{t}]
    &=
    \frac
    {
        \sigma_\Vect{t} \, 
        \alpha_{(0,t_2)}
        \sqrt{\varphi_{(0,t_2)}}
    }
    {
        \vartheta_\Vect{t}
        \sqrt{- 2 \pi i}
        \sqrt
        {
            k_2'(\Vect{t})
        }
    }
    \, \Upsilon_\Vect{t}^\rho[\Vect{q}(\Vect{t})]
    \nonumber\\
    &\times
    \exp
    \left\{
        i t_1 \frac{1 + \vartheta_\Vect{t}^2}{2}
        - \frac{i}{2} \beta_{(0, t_2)}^\rho [\Vect{q}(0, t_2)]
    \right\}
    .
    \label{eq:cuspPSIt}
\end{align}

\noindent Note that $k_2'(\Vect{t})$ can change sign, meaning $\sigma_\Vect{t} \neq 1$. However, we do not need to explicitly compute $\sigma_\Vect{t}$ since it will be removed by matching to initial conditions. 

After making a slow-envelope approximation, a quartic polynomial (normal form) can be fit to an implicit Taylor expansion of $\Theta_\Vect{t}\left[Q_{\Vect{t}, 1}(\Vect{t}), Q_{\Vect{t},2}(\Vect{t}) - \epsilon_\rho \right]$ to ultimately yield
\begin{align}
    \Upsilon_\Vect{t}^\rho[\Vect{q}(\Vect{t})]
    &\approx
    \vartheta_\Vect{t} 
    \left| 
        \frac{f}{\aber}
    \right|^{1/4}
    \sqrt{
        \frac{- 2 k_2'(\Vect{t})}{t_1}
    }
    \exp
    \left(
        i \frac{j_\Vect{t} + f - t_1}{4 t_1 } f t_2^2
    \right)
    \nonumber\\
    &\hspace{4mm}\times
    \int_{\cont{t_2}} \dd \varepsilon \,
    \exp
    \left(
        i  y_\Vect{t}
        \varepsilon
        + i x_\Vect{t}
        \varepsilon^2
        + i \varepsilon^4
    \right)
    ,
    \label{eq:cuspUPSILON}
\end{align}

\noindent where we have defined
\begin{equation}
    x_\Vect{t} \doteq
    \left| 
        \frac{ f}{ \aber}
    \right|^{1/2}
    \frac{
        f - q_1(\Vect{t})
    }{q_1(\Vect{t})}
    , \quad
    y_\Vect{t} \doteq
    \left| 
        \frac{4 f^3}{\aber} 
    \right|^{1/4}
    \frac{
        q_2(\Vect{t})
    }{q_1(\Vect{t})}
    ,
\end{equation}

\noindent and $\cont{t_2}$ is the steepest-descent contour through the saddlepoint $\varepsilon = - t_2 |f \aber|^{1/4}/\sqrt{2}$.

Our assumption $\aber < 0$ implies that $\Vect{\tau}(\Vect{q})$ is single-valued along the initial surface. Thus, \Eq{eq:cuspPSIt} yields
\begin{align}
    \hspace{-2mm}\psi\left[ \Vect{q}(0, t_2) \right]
    &=
    \sigma_{t_2} \, 
    \alpha_{(0,t_2)}
    \sqrt{
        \frac{ \varphi_{(0,t_2)} }{f }
    }
    \nonumber\\
    &\times
    \exp
    \left\{
        i \frac{1 - s_\Vect{t} }{2} \pi
        - \frac{i}{2} \beta_{(0, t_2)}^\rho \left[\Vect{q}(0, t_2) \right]
    \right\}
    ,
\end{align}

\noindent where we have defined $s_\Vect{t} \doteq \textrm{sgn}[k_2'(\Vect{t})]$ and have evaluated \Eq{eq:cuspUPSILON} in the GO limit, since $f \gg 1$ implies that the initial surface lies sufficiently far from the caustic. Thus, the initial conditions are satisfied by the choice
\begin{align}
    \alpha_{(0, t_2)} 
    &= 
    \frac
    {
        \sqrt{2 \pi}
    }
    { 
        \sigma_{t_2} 
        \sqrt{ \varphi_{(0,t_2)} } 
    }
    \exp
    \left\{
        \frac{i}{2} \beta_{(0, t_2)}^\rho \left[\Vect{q}(0, t_2) \right]
        - \frac{i}{2} f t_2^2
        \right.\nonumber\\
        &\left.\hspace{28mm}
        - \frac{i \aber}{4} f t_2^4
        + i \frac{ 2 s_\Vect{t} - 1}{4} \pi
    \right\}
    .
\end{align} 

\noindent Equation \eq{eq:MGOpsi} therefore yields
\begin{align}
    \psi(\Vect{q}) 
    &=
    \left|
        \frac{4 f}{\aber q_1^2}
    \right|^{1/4}
    \, 
    \exp
    \left(
        i q_1
        + i \frac{q_2^2}{2 q_1}
    \right)
    \nonumber\\
    &\times
    \sum_{t_2 \in \tau_2(\Vect{q})}
    \int_{\cont{t_2}} \dd \varepsilon \,
    \exp
    \left(
        i 
        y_\Vect{t}
        \varepsilon
        + i 
        x_\Vect{t}
        \varepsilon^2
        + i \varepsilon^4
    \right)
    ,
    \label{eq:cuspPSIsum}
\end{align}

\noindent where the sum is over all real saddlepoint contributions.

Let us recall the Pearcey function~\cite{Paris91}, defined as
\begin{equation}
    \pearcey(x, y) \doteq 
    \int_{-\infty}^\infty \dd s \, \exp \left(i y s + i x s^2 + i s^4 \right)
    .
\end{equation}

\noindent Then, when $\Delta(\Vect{q}) \widetilde{\Delta}(\Vect{q}) \ge 0$, where
\begin{equation}
    \widetilde{\Delta}(\Vect{q}) \doteq 
    2 \left(
        \frac{f - q_1}{|\aber| q_1}
    \right)^3
    + 27 (5 - \sqrt{27})
    \left(
        \frac{q_2}{|\aber| q_1}
    \right)^2
    \, ,
\end{equation}

\noindent the summation in \Eq{eq:cuspPSIsum} is simplified as
\begin{equation}
    \hspace{-5mm}\sum_{t_2 \in \tau_2(\Vect{q})}
    \int_{\cont{t_2}} \dd \varepsilon \,
    \exp
    \left(
        i 
        y_\Vect{t}
        \varepsilon
        + i 
        x_\Vect{t}
        \varepsilon^2
        + i \varepsilon^4
    \right)
    = \pearcey(x_\Vect{t},y_\Vect{t}) 
    .
\end{equation}

\noindent When $\Delta(\Vect{q}) \widetilde{\Delta}(\Vect{q}) < 0$ (the caustic shadow), $\pearcey(x,y)$ contains an additional contribution from one of the two complex saddlepoints~\cite{Wright80}, which is not included in \Eq{eq:cuspPSIsum}. It does not seem possible to isolate the real saddlepoint contribution to $\pearcey(x,y)$ using a complex rotation as done in \Ref{Lopez20a} for $\airyA(x)$. Nevertheless, the shadow contribution is asymptotically subdominant, so within the MGO accuracy we can include it such that \Eq{eq:cuspPSIsum} can be universally expressed through the Pearcey function as
\begin{align}
    \psi(\Vect{q})
    &=
    \left|
        \frac{4 f}{\aber q_1^2}
    \right|^{1/4}
    \, 
    \exp
    \left(
        i q_1
        + i \frac{q_2^2}{2 q_1}
    \right)
    \nonumber\\
    &\hspace{4mm}\times
    \pearcey
    \left(
        \left| 
            \frac{ f}{ \aber}
        \right|^{1/2}
        \frac{
            f - q_1
        }{q_1}
        , 
        \left| 
            \frac{4 f^3}{\aber} 
        \right|^{1/4}
        \frac{
            q_2
        }{q_1}
    \right)
    .
    \label{eq:cuspMGO}
\end{align}

\begin{figure}
    \centering
    \includegraphics[width=\linewidth,trim={2mm 4mm 4mm 3mm},clip]{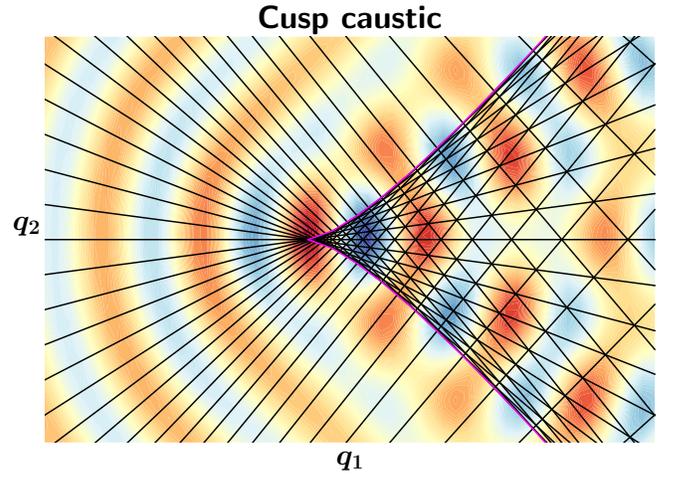}
    \caption{Contour plot showing the real part of the MGO solution \eq{eq:cuspMGO}, with $\aber = -4/f$, near a cusp caustic located at $q_2 = \pm \sqrt{ 4(f-q_1)^3 / 27\aber q_1 }$ (magenta). The ray trajectories $( q_1(\Vect{\tau}), q_2(\Vect{\tau}) )$ are shown as black lines. Note that the field remains finite along the caustic.}
    \label{fig:cusp}
\end{figure}

\noindent As readily verified, \Eq{eq:cuspMGO} also happens to be the exact solution of \Eq{eq:cuspEQ} for the initial condition \eq{eq:cuspIC}. This solution is illustrated in \Fig{fig:cusp} for $\aber = - 4/f$.

\section{Conclusion}
\label{sec:conclusion}

Metaplectic geometrical optics, or MGO, has recently been developed to accurately model caustics by integrating field equations over GO rays. However, as originally formulated in \Ref{Lopez20a}, MGO fails to describe what we call quasiuniform ray patterns ($\det \Mat{B}_\Vect{t} = 0$ over a finite domain). Here, we extend MGO so that the new theory can be applied to any ray pattern in both uniform and nonuniform media. To aid practical implementations, we provide two equivalent representations of the new MGO using either singular value decompositions or Gaussian states. 

We demonstrate MGO analytically in three examples, namely, a plane wave propagating in uniform media (no caustic), a plane wave incident on an isolated cutoff (Airy-type fold caustic), and an imperfectly focused plane wave in vacuum (Pearcey-type cusp caustic). In all examples, MGO provides an accurate representation of the exact solution that remains finite at the caustics, unlike traditional GO. Yet, as the final example shows, MGO does not always accurately model caustic shadows. Further extensions of MGO to incorporate caustic shadow fields will be investigated in future research. In future work we shall also explore how to include polarization dynamics for modeling vector waves with MGO, including spin--orbit coupling~\cite{Dodin19,Bliokh04a,Bliokh04b,Ruiz17a,Ruiz15a,Ruiz15b,Ruiz15d,Ruiz17t,Oancea20,Bliokh15}, mode conversion~\cite{Dodin19,Yanagihara19a,Yanagihara19b,Dodin17c}, and optical anisotropy~\cite{Ruiz17a,Kravtsov96,Kravtsov07,Bliokh07}.


\section*{Acknowledgements}

The work was supported by the U.S. DOE through Contract No.~DE-AC02-09CH11466.


\appendix


\section{Symplectic matrices in the SVD basis}
\label{app:Matrices}

As is well-known, the symplectic criterion \eq{eq:symplec} for $\Mat{S}$ implies that $\Mat{A}$, $\Mat{B}$, $\Mat{C}$, and $\Mat{D}$ satisfy~\cite{Lopez19a,Luneburg64}
\begin{subequations}
    \label{eq:symplecABCD}
    \begin{align}
        \label{eq:symplec1}
        \Mat{A} \Mat{B}^\intercal 
        - \Mat{B} \Mat{A}^\intercal 
        = \OMat{N} 
        , \\
        \label{eq:symplec2}
        \Mat{B}^\intercal \Mat{D} 
        - \Mat{D}^\intercal \Mat{B} 
        = \OMat{N}
        , \\
        \label{eq:symplec3}
        \Mat{A} \Mat{D}^\intercal 
        - \Mat{B} \Mat{C}^\intercal 
        = \IMat{N}
        , \\
        \label{eq:symplec4}
        \Mat{A}^\intercal \Mat{D} 
        - \Mat{C}^\intercal \Mat{B} 
        = \IMat{N}
        , \\
        \label{eq:symplec5}
        \Mat{C}^\intercal \Mat{A} - \Mat{A}^\intercal \Mat{C} = \OMat{N}
        , \\
        \label{eq:symplec6}
        \Mat{D} \Mat{C}^\intercal - \Mat{C} \Mat{D}^\intercal = \OMat{N} .
    \end{align}
\end{subequations}

\noindent The orthogonality of $\Mat{L}$ and $\Mat{R}$ means that $\widetilde{\Mat{A}}$, $\widetilde{\Mat{B}}$, $\widetilde{\Mat{C}}$, and $\widetilde{\Mat{D}}$ also satisfy \Eqs{eq:symplecABCD}. After using the parameterization
\begin{equation}
    \widetilde{\Mat{A}} = 
    \begin{pmatrix}
        \Mat{a}_{\rho \rho} & \Mat{a}_{\rho \varsigma} \\[1mm]
        \Mat{a}_{\varsigma \rho} & \Mat{a}_{\varsigma \varsigma}
    \end{pmatrix}
\end{equation}

\noindent (where each block $\Mat{a}_{mn}$ is a matrix of size $m \times n$), \Eq{eq:symplec1} reads
\begin{equation}
    \begin{pmatrix}
        \Mat{a}_{\rho \rho} \Mat{\Lambda}_{\rho \rho} - \left(\Mat{a}_{\rho \rho} \Mat{\Lambda}_{\rho \rho} \right)^\intercal 
        & -\left( \Mat{a}_{\varsigma \rho} \Mat{\Lambda}_{\rho \rho} \right)^\intercal \\[1mm]
        \Mat{a}_{\varsigma \rho} \Mat{\Lambda}_{\rho \rho} & \OMat{\varsigma \varsigma}
    \end{pmatrix}
    = \OMat{N}
    .
\end{equation}

\noindent Since $\Mat{\Lambda}_{\rho \rho} \neq \OMat{\rho \rho}$, this implies that
\begin{equation}
    \Mat{a}_{\varsigma \rho} = \OMat{\varsigma \rho}
    ,
    \quad
    \Mat{a}_{\rho \rho} \Mat{\Lambda}_{\rho \rho}
    = 
    \left(
        \Mat{a}_{\rho \rho} \Mat{\Lambda}_{\rho \rho}
    \right)^\intercal 
    .
    \label{eq:aSUBsymm}
\end{equation}

\noindent Similarly, after using the parameterization
\begin{equation}
    \widetilde{\Mat{D}} = 
    \begin{pmatrix}
        \Mat{d}_{\rho \rho} & \Mat{d}_{\rho \varsigma} \\[1mm]
        \Mat{d}_{\varsigma \rho} & \Mat{d}_{\varsigma \varsigma}
    \end{pmatrix}
    ,
\end{equation}

\noindent \Eq{eq:symplec2} becomes
\begin{equation}
    \begin{pmatrix}
        \Mat{\Lambda}_{\rho \rho} \Mat{d}_{\rho \rho} - \left( \Mat{\Lambda}_{\rho \rho} \Mat{d}_{\rho \rho} \right)^\intercal
        & \Mat{\Lambda}_{\rho \rho} \Mat{d}_{\rho \varsigma}
        \\[1mm]
        - \left(\Mat{\Lambda}_{\rho \rho} \Mat{d}_{\rho \varsigma} \right)^\intercal
        &
        \OMat{\varsigma \varsigma}
    \end{pmatrix}
    = 
    \OMat{N} .
\end{equation}

\noindent This yields analogous constraints on $\widetilde{\Mat{D}}$, namely
\begin{equation}
    \Mat{d}_{\rho \varsigma} = \OMat{\rho \varsigma}
    ,
    \quad
    \Mat{\Lambda}_{\rho \rho} \Mat{d}_{\rho \rho} 
    = 
    \left( 
        \Mat{\Lambda}_{\rho \rho} \Mat{d}_{\rho \rho} 
    \right)^\intercal .
    \label{eq:dSUBsymm}
\end{equation}

Next, after using the parameterization
\begin{equation}
    \widetilde{\Mat{C}} = 
    \begin{pmatrix}
        \Mat{c}_{\rho \rho} & \Mat{c}_{\rho \varsigma} \\[1mm]
        \Mat{c}_{\varsigma \rho} & \Mat{c}_{\varsigma \varsigma}
    \end{pmatrix}
    ,
\end{equation}

\noindent \Eq{eq:symplec3} becomes
\begin{equation}
    \begin{pmatrix}
        \Mat{a}_{\rho \rho} \Mat{d}_{\rho \rho}^\intercal - \Mat{\Lambda}_{\rho \rho} \Mat{c}_{\rho \rho}^\intercal
        &
        \Mat{a}_{\rho \rho} \Mat{d}_{\varsigma \rho}^\intercal + \Mat{a}_{\rho \varsigma} \Mat{d}_{\varsigma \varsigma}^\intercal - \Mat{\Lambda}_{\rho \rho} \Mat{c}_{\varsigma \rho}^\intercal
        \\[1mm]
        \OMat{\varsigma \rho}
        &
        \Mat{a}_{\varsigma \varsigma} \Mat{d}_{\varsigma \varsigma}^\intercal
    \end{pmatrix}
    =
    \IMat{N} .
\end{equation}

\noindent Since the matrix inverse is unique, we therefore obtain
\begin{equation}
    \Mat{d}_{\varsigma \varsigma} 
    =
    \Mat{a}_{\varsigma \varsigma}^{-\intercal}
    \label{eq:dINV}
    ,
\end{equation}

\noindent which means that $\Mat{a}_{\varsigma \varsigma}$ is invertible. Since $\Mat{\Lambda}_{\rho \rho}$ is invertible, we also obtain
\begin{subequations}
    \begin{align}
        \Mat{c}_{\rho \rho} 
        &= \Mat{d}_{\rho \rho} \Mat{a}_{\rho \rho}^\intercal \Mat{\Lambda}_{\rho \rho}^{-1} 
        - \Mat{\Lambda}_{\rho \rho}^{-1}
        ,
        \\
        \Mat{c}_{\varsigma \rho} 
        &= \Mat{d}_{\varsigma \rho} \Mat{a}_{\rho \rho}^\intercal \Mat{\Lambda}_{\rho \rho}^{-1} 
        + \Mat{a}_{\varsigma \varsigma}^{-\intercal} \Mat{a}_{\rho \varsigma}^\intercal \Mat{\Lambda}_{\rho \rho}^{-1}
        ,
    \end{align}
\end{subequations}

\noindent where we have used \Eq{eq:dINV}. Consequently, \Eq{eq:symplec4} is greatly simplified; it reads
\begin{equation}
    \begin{pmatrix}
        \IMat{\rho}
        &
        \OMat{\rho \varsigma}
        \\[1mm]
        \Mat{a}_{\rho \varsigma}^\intercal \Mat{d}_{\rho \rho}
        +
        \Mat{a}_{\varsigma \varsigma}^\intercal \Mat{d}_{\varsigma \rho}
        -
        \Mat{c}_{\rho \varsigma}^\intercal \Mat{\Lambda}_{\rho \rho}
        &
        \IMat{\varsigma}
    \end{pmatrix}
    =
    \IMat{N} .
\end{equation}

\noindent We therefore obtain
\begin{equation}
    \Mat{c}_{\rho \varsigma} = \Mat{\Lambda}_{\rho \rho}^{-1} \Mat{d}_{\rho \rho}^\intercal \Mat{a}_{\rho \varsigma} + \Mat{\Lambda}_{\rho \rho}^{-1} \Mat{d}_{\varsigma \rho}^\intercal \Mat{a}_{\varsigma \varsigma}
    .
\end{equation}

\noindent Finally, since $\Mat{a}_{\varsigma \varsigma}$ is invertible, both \Eq{eq:symplec5} and \Eq{eq:symplec6} take the form
\begin{equation}
    \begin{pmatrix}
        \OMat{\rho \rho}
        &
        \OMat{\rho \varsigma} 
        \\
        \OMat{\varsigma \rho}
        &
        \Mat{n}
        -
        \Mat{n}^\intercal
    \end{pmatrix}
    =
    \OMat{N} ,
\end{equation}

\noindent where 
\begin{equation}
    \Mat{n}
    \doteq
    \Mat{a}_{\varsigma \varsigma}^\intercal \Mat{d}_{\varsigma \rho} \Mat{\Lambda}_{\rho \rho}^{-1} \Mat{a}_{\rho \varsigma}
    - 
    \Mat{a}_{\varsigma \varsigma}^\intercal \Mat{c}_{\varsigma \varsigma} .
\end{equation}

\noindent We therefore require $\Mat{c}_{\varsigma \varsigma}$ to satisfy
\begin{equation}
    \Mat{a}_{\varsigma \varsigma}^\intercal \Mat{d}_{\varsigma \rho} \Mat{\Lambda}_{\rho \rho}^{-1} \Mat{a}_{\rho \varsigma}
    - 
    \Mat{a}_{\varsigma \varsigma}^\intercal \Mat{c}_{\varsigma \varsigma}
    =
    \left(
        \Mat{a}_{\varsigma \varsigma}^\intercal \Mat{d}_{\varsigma \rho} \Mat{\Lambda}_{\rho \rho}^{-1} \Mat{a}_{\rho \varsigma}
        - 
        \Mat{a}_{\varsigma \varsigma}^\intercal \Mat{c}_{\varsigma \varsigma}
    \right)^\intercal .
    \label{eq:cSUBsymm}
\end{equation}


\section{Derivation of equation \eq{eq:MTepsilon}}
\label{app:MTphase}

Here, we derive \Eq{eq:MTepsilon} as the limit of $M(\Vect{Q}, \Vect{q}; \Mat{S}_\varepsilon)$ at $\varepsilon \to 0$. First, we obtain, using \Eq{eq:METkern},
\begin{align}
    M(\Vect{Q},\Vect{q}; \Mat{S}_\varepsilon)
    &=
    \frac
    {
        \sigma \,
        \exp
        \left[ 
            i \widetilde{G}(\Vect{q}, \Vect{Q})
        \right]
    }
    {
        (2 \pi i)^{N/2}
        \sqrt
        {
            \det
            \left(
                \Mat{B} + \varepsilon \Mat{A}
            \right)
        }
    }
    ,
\end{align}

\noindent where we have defined
\begin{align}
    \widetilde{G}(\Vect{q}, \Vect{Q})
    &\doteq
    \frac{1}{2}
    \left(
        \Mat{L}^\intercal \Vect{Q}
    \right)^\intercal
    \left(
        \widetilde{\Mat{D}} + \varepsilon \widetilde{\Mat{C}}
    \right)
    \left(
        \widetilde{\Mat{B}} + \varepsilon \widetilde{\Mat{A}} 
    \right)^{-1}
    \Mat{L}^\intercal
    \Vect{Q}
    \nonumber\\
    &-
    \left(
        \Mat{R}^\intercal \Vect{q}
    \right)^\intercal
    \left(
        \widetilde{\Mat{B}} + \varepsilon \widetilde{\Mat{A}}
    \right)^{-1}
    \left(
        \Mat{L}^\intercal \Vect{Q}
        +\frac{1}{2} \widetilde{\Mat{A}} \,
    \Mat{R}^\intercal \Vect{q}
    \right)
\end{align}

\noindent and used the unitarity of $\Mat{L}$ and $\Mat{R}$. Next, we must approximate the matrix inverse term to leading order in $\varepsilon$. Let us adopt the parameterization
\begin{equation}
    \left(
        \widetilde{\Mat{B}}
        + \varepsilon \widetilde{\Mat{A}}
    \right)^{-1}
    =
    \begin{pmatrix}
        \Mat{m}_{\rho \rho} & \Mat{m}_{\rho \varsigma} \\
        \Mat{m}_{\varsigma \rho} & \Mat{m}_{\varsigma \varsigma}
    \end{pmatrix}
    .
\end{equation}

\noindent Then, since 
\begin{equation}
    \begin{pmatrix}
        \Mat{m}_{\rho \rho}
        \left(
            \Mat{\Lambda}_{\rho \rho} 
            + \varepsilon \Mat{a}_{\rho \rho}
        \right)
        &
        \varepsilon
        \left(
            \Mat{m}_{\rho \rho} \Mat{a}_{\rho \varsigma} 
            + \Mat{m}_{\rho \varsigma} \Mat{a}_{\varsigma \varsigma}
        \right)
        \\[1mm]
        \Mat{m}_{\varsigma \rho} 
        \left(
            \Mat{\Lambda}_{\rho \rho} 
            + \varepsilon \Mat{a}_{\rho \rho}
        \right)
        &
        \varepsilon
        \left(
            \Mat{m}_{\varsigma \rho} \Mat{a}_{\rho \varsigma}
            + \Mat{m}_{\varsigma \varsigma} \Mat{a}_{\varsigma \varsigma}
        \right)
    \end{pmatrix}
    = \IMat{N}
    ,
\end{equation}

\noindent we require
\begin{subequations}
    \label{eq:mEQs}
    \begin{align}
        \Mat{m}_{\varsigma \rho} 
        \left(
            \Mat{\Lambda}_{\rho \rho} 
            + \varepsilon \Mat{a}_{\rho \rho}
        \right)
        &= \OMat{\varsigma \rho}
        , 
        \\
        \varepsilon
        \left(
            \Mat{m}_{\rho \rho} \Mat{a}_{\rho \varsigma} 
            + \Mat{m}_{\rho \varsigma} \Mat{a}_{\varsigma \varsigma}
        \right)
        &= \OMat{\rho \varsigma}
        ,
        \\
         \varepsilon
        \left(
            \Mat{m}_{\varsigma \rho} \Mat{a}_{\rho \varsigma}
            + \Mat{m}_{\varsigma \varsigma} \Mat{a}_{\varsigma \varsigma}
        \right)
        &= \IMat{\varsigma}
        ,
        \\
        \Mat{m}_{\rho \rho}
        \left(
            \Mat{\Lambda}_{\rho \rho} 
            + \varepsilon \Mat{a}_{\rho \rho}
        \right)
        &= \IMat{\rho}
        .
    \end{align}
\end{subequations}

\noindent Solving \Eqs{eq:mEQs} in sequence yields
\begin{subequations}
    \begin{align}
        \Mat{m}_{\varsigma \rho} 
        &= \OMat{\varsigma \rho}
        ,
        \quad
        \Mat{m}_{\rho \varsigma} 
        = - \Mat{m}_{\rho \rho} \Mat{a}_{\rho \varsigma} \Mat{a}_{\varsigma \varsigma}^{-1}
        ,
        \\
        \Mat{m}_{\varsigma \varsigma}
        &= \varepsilon^{-1} \Mat{a}_{\varsigma \varsigma}^{-1}
        ,
        \quad
        \Mat{m}_{\rho \rho}
        = (\Mat{\Lambda}_{\rho \rho} + \varepsilon \Mat{a}_{\rho \rho})^{-1}
        \approx \Mat{\Lambda}_{\rho \rho}^{-1}
        ,
    \end{align}
\end{subequations}

\noindent where we have used the fact that $\Mat{a}_{\varsigma \varsigma}$ is invertible. Hence, we obtain the following leading-order approximations:
\begin{align}
    &\left(
        \widetilde{\Mat{D}} 
        + \varepsilon \widetilde{\Mat{C}}
    \right)
    \left(
        \widetilde{\Mat{B}} 
        + \varepsilon \widetilde{\Mat{A}}
    \right)^{-1}
    \nonumber\\
    &\approx
    \begin{pmatrix}
        \Mat{d}_{\rho \rho} \Mat{\Lambda}_{\rho \rho}^{-1}
        &
        \Mat{\Lambda}_{\rho \rho}^{-1} \Mat{d}_{\varsigma \rho}^\intercal
        \\[1mm]
        \Mat{d}_{\varsigma \rho} \Mat{\Lambda}_{\rho \rho}^{-1}
        &
        \varepsilon^{-1} \Mat{a}_{\varsigma \varsigma}^{-\intercal} \Mat{a}_{\varsigma \varsigma}^{-1} + \Mat{c}_{\varsigma \varsigma} \Mat{a}_{\varsigma \varsigma}^{-1} - \Mat{d}_{\varsigma \rho} \Mat{\Lambda}_{\rho \rho}^{-1} \Mat{a}_{\rho \varsigma} \Mat{a}_{\varsigma \varsigma}^{-1}
    \end{pmatrix}
    ,
    \\
    &\left(
        \widetilde{\Mat{B}} 
        + \varepsilon \widetilde{\Mat{A}}
    \right)^{-1}
    \widetilde{\Mat{A}}
    \approx
    \begin{pmatrix}
        \Mat{\Lambda}_{\rho \rho}^{-1} \Mat{a}_{\rho \rho}
        &
        \OMat{\rho \varsigma}
        \\[1mm]
        \OMat{\varsigma \rho}
        &
        \varepsilon^{-1} \IMat{\varsigma}
    \end{pmatrix}
    .
\end{align}

\noindent Upon introducing the vector decompositions
\begin{equation}
    \Mat{R}^\intercal\Vect{x}
    =
    \begin{pmatrix}
        \Vect{x}_\rho \\[1mm]
        \Vect{x}_\varsigma
    \end{pmatrix}
    , \quad
    \Mat{L}^\intercal \Vect{y}
    =
    \begin{pmatrix}
        \Vect{y}_\rho \\[1mm]
        \Vect{y}_\varsigma
    \end{pmatrix}
\end{equation}

\noindent (where all subvectors $\Vect{\nu}_n$ are size $n \times 1$), we obtain
\begin{align}
    \widetilde{G}(\Vect{q}, \Vect{Q})
    &\approx
    \frac{1}{2\varepsilon} 
    \left|
        \Vect{q}_\varsigma 
        - \Mat{a}_{\varsigma \varsigma}^{-1} \Vect{Q}_\varsigma 
    \right|^2
    + \frac{1}{2} \Vect{q}_\rho^\intercal \, \Mat{\Lambda}_{\rho \rho}^{-1} \Mat{a}_{\rho \rho} \, \Vect{q}_\rho
    \nonumber\\
    &\hspace{4mm}
    - \Vect{q}_\rho^\intercal \, \Mat{M}_1 \, \Mat{L}^\intercal \Vect{Q}
    + \frac{1}{2} \Vect{Q}^\intercal \, \Mat{L} \Mat{M}_2 \Mat{L}^\intercal \, \Vect{Q}
    ,
\end{align}

\noindent where the matrices $\Mat{M}_1$ and $\Mat{M}_2$ are defined in \Eqs{eq:M12}.


\section{Metaplectic transforms in the mixed basis of configuration and coherent states}
\label{app:Gauss}

Here, we derive \Eq{eq:gaussMT} using Gaussian coherent states. To help with the presentation, in this section, we employ the bra-ket notation of quantum mechanics~\cite{Stoler81}. 

The Gaussian coherent states, denoted $\ket{\Stroke{\Vect{Z}}_0}$, are defined by their $\Vect{Q}$-space representations
\begin{equation}
    \braket{\Vect{Q}}{\Stroke{\Vect{Z}}_0}
    =
    \frac
    {
        \exp
        \left[
            - \frac{|\Vect{Q} - \Vect{Q}_0|^2}{2}
            + i \Vect{K}_0^\intercal 
            \left(
                \Vect{Q}
                - \frac{\Vect{Q}_0}{2}
            \right)
        \right] 
    }
    {
        \pi^{N/4}
    }
    ,
\end{equation}

\noindent where $\ket{\Vect{Q}}$ are the eigenstates of the $\Vect{Q}$-space position operator $\VectOp{Q}$, normalized as $\braket{\Vect{Q}}{\Vect{Q}'} = \delta(\Vect{Q} - \Vect{Q}')$. The parameters $\Vect{Q}_0$ and $\Vect{K}_0$ define the center of $\ket{\Stroke{\Vect{Z}}_0}$ in phase space, that is,
\begin{equation}
    \bra{\Stroke{\Vect{Z}}_0}
    \VectOp{Q}
    \ket{\Stroke{\Vect{Z}}_0}
    = \Vect{Q}_0
    ,
    \quad
    \bra{\Stroke{\Vect{Z}}_0}
    \VectOp{P}
    \ket{\Stroke{\Vect{Z}}_0}
    = \Vect{K}_0
    ,
\end{equation}

\noindent where $\VectOp{P}$ is the $\Vect{Q}$-space momentum operator. The states $\ket{\Stroke{\Vect{Z}}_0}$ also satisfy a completeness relation of the form
\begin{equation}
    \IdentOp = \int \frac{\dd \Vect{Q}_0 \dd \Vect{K}_0}{(2\pi)^N} \, \ket{\Stroke{\Vect{Z}}_0}
    \bra{\Stroke{\Vect{Z}}_0} ,
\end{equation}

\noindent where $\IdentOp$ denotes the identity operator.

Let us denote $\psi_\Vect{t}(\Vect{q})$ by $\braket{\Vect{q}}{\psi_\Vect{t}}$ and $\Psi_\Vect{t}(\Vect{Q})$ by $\braket{\Vect{Q}}{\psi_\Vect{t}}$. Then, the inverse MT of \Eq{eq:metINV} can be written as
\begin{equation}
    \psi_\Vect{t}(\Vect{q}) 
    =
    \int \dd \Vect{Q} \,
    \braket{\Vect{q}}{\Vect{Q}}
    \braket{\Vect{Q}}{\psi_\Vect{t}}
    =
    \int \dd \Vect{Q} \,
    \braket{\Vect{q}}{\Vect{Q}}
    \Psi(\Vect{Q}) ,
\end{equation}

\noindent where we have used the completeness of $\ket{\Vect{Q}}$, that is,
\begin{equation}
    \IdentOp = \int \dd \Vect{Q} \,
    \ket{\Vect{Q}}
    \bra{\Vect{Q}}
    .
\end{equation}

\noindent The matrix element $\braket{\Vect{q}}{\Vect{Q}}$ is the inverse MT kernel. Direct computation of $\braket{\Vect{q}}{\Vect{Q}}$ will yield \Eq{eq:invMETkern}, which is delta-shaped when $\det \Mat{B}_\Vect{t} = 0$. However, by introducing the normalizable states $\ket{\Stroke{\Vect{Z}}_0}$ as
\begin{equation}
    \braket{\Vect{q}}{\Vect{Q}}
    =
    \int \frac{\dd \Vect{Q}_0 \dd \Vect{K}_0}{(2\pi)^N} \,
    \braket{\Vect{q}}{\Stroke{\Vect{Z}}_0}
    \braket{\Stroke{\Vect{Z}}_0}{\Vect{Q}},
    \label{eq:gaussINT}
\end{equation}

\noindent a nonsingular representation of the MT can be obtained.

Indeed, as shown in \Ref{Littlejohn86a}, 
\begin{align}
    &\braket{\Vect{q}}{\Stroke{\Vect{Z}}_0}
    =
    \frac
    {
        \sigma_\Vect{t}
        \,
        \exp
        \left[
            -\frac{1}{2} \Vect{q}^\intercal 
        \left(
            \Mat{D}_\Vect{t}
            - i \Mat{B}_\Vect{t}
        \right)^{-1}
        \left(
            \Mat{A}_\Vect{t}
            + i \Mat{C}_\Vect{t}
        \right) \Vect{q}
        \right]
    }
    {
        \pi^{N/4} 
        \sqrt{ \det(\Mat{D}_\Vect{t} - i \Mat{B}_\Vect{t} ) }
    }
    \nonumber\\
    &\times
    \exp
    \left[
        \left(
            \Vect{q}
            - \frac{i}{2} \Mat{B}_\Vect{t}^\intercal \Vect{\zeta}
        \right)^\intercal
        \left(
            \Mat{D}_\Vect{t}
            - i \Mat{B}_\Vect{t}
        \right)^{-1}
        \Vect{\zeta}
        - \frac{1}{2} \Vect{Q}_0^\intercal \Vect{\zeta}
    \right]
    ,
\end{align}

\noindent where we have introduced the complex vector $\Vect{\zeta} \doteq \Vect{Q}_0 + i \Vect{K}_0$. Importantly, the complex matrix $\Mat{D}_\Vect{t} - i \Mat{B}_\Vect{t}$ is always invertible~\cite{Littlejohn87}. Then, upon using $\braket{\Stroke{\Vect{Z}}_0}{\Vect{Q}} = ( \braket{\Vect{Q}}{\Stroke{\Vect{Z}}_0} )^*$, one computes
\begin{align}
    &\braket{\Vect{q}}{\Stroke{\Vect{Z}}_0}
    \braket{\Stroke{\Vect{Z}}_0}{\Vect{Q}}
    =
    \frac
    {
        \sigma_\Vect{t}
        \exp
        \left[
            - \frac{1}{2} \Vect{q}^\intercal 
            \left(
                \Mat{D}_\Vect{t}
                - i \Mat{B}_\Vect{t}
            \right)^{-1} 
            \left(
                \Mat{A}_\Vect{t}
                + i \Mat{C}_\Vect{t}
            \right)
            \Vect{q}
        \right]
    }
    {
        \pi^{N/2} 
        \sqrt{ \det(\Mat{D}_\Vect{t} - i \Mat{B}_\Vect{t} ) }
    }
    \nonumber\\
    &\times
    \exp
    \left[
        \left(
            \Vect{q}
            - \frac{i}{2} \Mat{B}_\Vect{t}^\intercal \Vect{\zeta}
        \right)^\intercal
        \left(
            \Mat{D}_\Vect{t}
            - i \Mat{B}_\Vect{t}
        \right)^{-1} \Vect{\zeta}
        - |\Vect{Q}_0|^2
        \right.\nonumber\\
        &\left.\hspace{47mm}
        - \frac{|\Vect{Q}|^2}{2}
        + \Vect{Q}^\intercal \Vect{\zeta}^*
    \right]
    .
\end{align}

\noindent Finally, we integrate over $\Vect{Q}_0$ in \Eq{eq:gaussINT} to obtain \Eq{eq:gaussMT}. Note that $\Vect{K}_0$ cannot be integrated over without inverting $\Mat{B}_\Vect{t}$.

\bibliography{Biblio.bib}
\bibliographystyle{apsrev4-1}
\end{document}